\documentclass[11pt]{article}
\usepackage{latexsym,amssymb,amstext,amsmath}
\usepackage{hyperref}
\allowdisplaybreaks

\newcommand{\be}{\begin{equation}}
\newcommand{\ee}{\end{equation}}
\newcommand{\bea}{\begin{eqnarray}}
\newcommand{\eea}{\end{eqnarray}}

\begin{document}
\begin{titlepage}

\hfill LMU-ASC 73/15\\[-3ex]

\vskip 6mm

\begin{center}
\vskip 6mm

{\LARGE \textbf{
Nernst branes with Lifshitz asymptotics in \\
\vskip 2mm
 ${\cal N}=2$ gauged supergravity }}
\vskip 8mm

\textbf{G.L.~Cardoso$^{a}$, M.~Haack$^{b}$ and S.~Nampuri$^{a}$}

\vskip 6mm
$^a${\em  Center for Mathematical Analysis, Geometry and Dynamical Systems,\\
  Instituto Superior T\'ecnico, Universidade de Lisboa,\\
  Av. Rovisco Pais, 1049-001 Lisboa, Portugal}\\[.2ex]
$^b${\em Arnold Sommerfeld Center for Theoretical Physics,\\
Ludwig-Maximilians-Universit\"at M\"unchen,\\
Theresienstrasse 37, 80333 M\"unchen, Germany} \\[.2ex]

\end{center}

\vskip .2in
\begin{center} {\bf ABSTRACT } \end{center}
\begin{quotation}\noindent 
We discuss two classes of non-supersymmetric interpolating solutions in ${\cal N}=2$, $D=4$ gauged supergravity, that flow
from either a $z=2$ Lifshitz geometry or a conformal AdS background to the near-horizon geometry 
of a Nernst brane. We obtain these solutions by constructing a $z=2$ supersymmetric Lifshitz solution in the STU model from a first-order
rewriting of the action, then lifting it up to a five-dimensional background and subsequently modifying this five-dimensional solution 
by a two-parameter family of deformations. Under reduction, these give  
four-dimensional non-supersymmetric Nernst brane solutions.
This is a step towards resolving the Lifshitz tidal force singularity in the context of ${\cal N}=2$ gauged supergravity 
and suggests an approach to 
encoding the Nernst brane in terms of the Schr\"odinger symmetry group of the holographically dual field theory.

\end{quotation}
\vfill
\today
\end{titlepage}
\eject

\section{Introduction}

One of the most far-ranging and conceptually incisive insights to emerge from string theory is the celebrated AdS/CFT correspondence \cite{Maldacena:1997re,Gubser:1998bc,Witten:1998qj}, which maps the Hilbert space and interactions of a lower dimensional 
conformal field theory to states and interactions in quantum gravity in asymptotically AdS backgrounds. This paves the way 
for a new template to conceive of and analyse the structure of quantum field theories grouped by their universality class.
Within this framework,
black holes offer a fecund laboratory to test and advance the conjectured holographic duality between gravity and gauge theories. 
Under this map, entropic black hole solutions in gravity map to thermal ensembles in the lower dimensional dual field theory.
The gauge-gravity duality, being a strong-weak coupling correspondence, potentially offers, via this map, a radically new approach to address
problems in ensembles of strongly coupled field theories in terms of the dynamics of classical fields in the 
dual black hole backgrounds. Adopting this promising 
program involves pursuing a two-fold strategy: firstly to identify the universality classes of field theories that are 
dual to existing black holes, and secondly and of unequivocal significance to quantum field theorists, to identify and construct 
black solutions, dual to states in quantum field theories which exhibit non-perturbative phenomena such as phase transitions, cf.\ 
\cite{Hartnoll:2009sz,McGreevy:2009xe,Hartnoll:2011fn} 
and references therein. An exciting research avenue that has opened up, in this spirit, focuses on modelling different states of 
particular field theories which 
can undergo quantum critical phase transitions by their holographic duals which consist of extremal black backgrounds. 

On the way to achieve this goal, solutions called Nernst branes were constructed in ${\cal N}=2 \; U(1)$  gauged supergravity theories 
in four dimensions \cite{Barisch:2011ui}. They were shown in \cite{Dempster:2015xqa} to be the extremal limits of well behaved black brane solutions which 
obey the third law of thermodynamics (i.e.\ their entropy density approaches zero for vanishing temperature), 
a prerequisite for modelling condensed matter systems. 
The extremal Nernst branes in \cite{Barisch:2011ui} interpolate between conformal AdS, and a conformal
hyperscale violating Lifshitz near-horizon geometry.
Embedding these solutions
as states in a field theory dual to conformal AdS was naively unfeasible, as the asymptotic symmetry group is not known for this background.
However, embedding Nernst branes in non-relativistic backgrounds such as Lifshitz backgrounds
offers a way to achieve this, as follows.
The Lifshitz spacetime is a non-relativistic background that preserves a sub-group of the Galilean symmetry group corresponding to broken boost symmetries \cite{Son:2007}, as opposed to
the full Poincar\'e group. Hence, the Lifshitz spacetime is invariant only under spatial rotations,
translations, and non-relativistic scaling. Further, it is not a smooth solution as it has
a tidal force singularity at its center
\cite{Kachru:2008yh,Hartnoll:2009sz,Charmousis:2010zz,Copsey:2010ya,Horowitz:2011gh}.
The only Lifshitz spacetime for which the asymptotic symmetry group
is known is the so-called $z=2$ Lifshitz spacetime. In this case, the symmetry algebra
governing the spectrum of the putative holographically dual field theory was identified as the Schr\"odinger symmetry group \cite{Hartong:2014pma}.\footnote{
The $z=2$ Lifshitz spacetime can be viewed as the Scherk-Schwarz reduction of a Poincar\'e-breaking
deformation of $AdS_5$ \cite{Blau:2009gd,
Balasubramanian:2010uk,Chemissany:2012du}. The Schr\"odinger asymptotic symmetry group
was established by analysing the effect of the deformation of AdS on its conformal symmetry
algebra \cite{Hartong:2014pma}.} Hence, constructing
Nernst branes in $z=2$ Lifshitz backgrounds opens up a way to obtain a
field theoretic understanding of
Nernst branes while simultaneously resolving the Lifshitz tidal force singularity.

\subsection{Calculational setup}
In classical supergravity, singular geometries such as Lifshitz space-times might, 
under deformation, flow to extremal black solutions that cloak the singularity or 
to smooth relativistic or non-relativistic BPS vacua. These vacua are attractors for the scalar flows and the resulting
flow equations are of first order, cf.\ \cite{Ferrara:1996dd,Ferrara:1996um,Ferrara:1997tw,Goldstein:2005hq} for the ungauged case. 
Such solutions have been extensively studied to obtain 
relativistic flows in  ${\cal N}=2, D=4$ gauged supergravity 
\cite{Cacciatori:2009iz,Dall'Agata:2010gj,Hristov:2010ri,Barisch:2011ui,Katmadas:2014faa}, wherein gravity is coupled to massless gauge fields 
of the vector multiplets. 
In this paper, we expand this analysis to advance the study of non-relativistic solution spaces.
Non-relativistic backgrounds are backgrounds with broken Poincar\'e symmetry.
One way to achieve this breaking is via a massive $U(1)$ gauge field \cite{Son:2008ye,Balasubramanian:2008dm}, and so we allow for  
the gauging of isometries of the quaternionic K\"ahler manifold that the hypermultiplet scalars are coordinates on. 
The resulting minimal coupling generates an effective Proca term in the Lagrangian. 
Non-relativistic supersymmetric solutions in  ${\cal N}=2, D=4$ gauged supergravity with hypermultiplets 
have been studied in \cite{Cassani:2011sv,Halmagyi:2011xh,Chimento:2015rra}.

The primary objective of our analysis in this note is to find a solution of supergravity that interpolates between an asymptotic $z=2$ Lifshitz geometry and another geometry that is regular and, in particular, does not suffer from any tidal force singularity. If this other geometry was a vacuum solution (for example $AdS_4$ or $AdS_2 \times \mathbb{R}^2$), such an interpolating solution would be a solitonic solution. In the dual field theory, it would correspond to an RG flow from a Lifshitz fixed point in the UV to a stable IR fixed point. Alternatively, the other geometry could be a regular non-vacuum solution. Given these two distinct possibilities motivates us to pursue two different strategies when examining the solution space of the low-energy gauged supergravity Lagrangian (and in particular that part of the solution space with Lifshitz asymptotics).

In a first attempt we search for solitonic solutions. These are expected to satisfy first-order flow equations resulting from a BPS rewriting of the action. Therefore, in sec.\ \ref{sec:rewriting} we perform a first-order rewriting 
of the Lagrangian (with purely electric gauging) to formulate first-order flow equations, 
which are in  perfect accord with those derived in \cite{Halmagyi:2013sla} by 
imposing supersymmetry conditions. In the case without
hypermultiplets a similar rewriting was performed in \cite{Barisch:2011ui}. However, given that in string theory 
compactifications with ${\cal N} = 2$ supersymmetry one always has at least the universal 
hypermultiplet, the inclusion of hypermultiplets is a crucial step towards making contact with actual string theory models and, thus, to 
embedding non-relativistic solutions into string theory. 
In addition, 
our first-order rewriting allows for a straightforward generalization that yields first-order flow equations for a particular class of non-supersymmetric solutions, as we discuss briefly at the end of 
sec.\ \ref{sec:nonsusy}. 

The solution space of these first-order flow equations is then shown in sec.\ \ref{sec:backgrounds} to support
$z=2$ Lifshitz geometries, which have a tidal force singularity. These solutions are written down for the STU model, whose cubic 
prepotential is more general than the $t^3$-model studied in \cite{Cassani:2011sv,Halmagyi:2011xh} in which Lifshitz solutions have been previously 
obtained. The charge and flux quantum numbers that support our Lifshitz solutions are found to be incompatible with the 
usual relativistic vacua, such as $AdS_4$ or $AdS_2 \times \mathbb{R}^2$, preventing us from constructing 
an interpolating solution between a Lifshitz and an AdS geometry. 

Not having found a solitonic solution with Lifshitz asymptotics, we change strategy according to the general twofold strategy outlined above, and we look for non-solitonic solutions interpolating between an asymptotic Lifshitz geometry and another regular solution in the center. This is done in sec.\ \ref{sec:interpolating}. 
We make use of the fact that the class of 
Lifshitz backgrounds found in sec.\ \ref{sec:backgrounds} can be obtained via a Scherk-Schwarz reduction of a Poincar\'e breaking background in five dimensions \cite{Chemissany:2012du}.
Next we study deformations of the five-dimensional line element that are consistent with the second order 
equations of motion of the five-dimensional theory, arriving at
a 2-parameter family of solutions which
was first obtained in \cite{Chemissany:2011mb}.
We then perform a Scherk-Schwarz reduction of these solutions 
to obtain two classes of four-dimensional interpolating solutions. One class indeed flows from
a $z=2$ Lifshitz background to the near-horizon geometry of the Nernst brane constructed in \cite{Barisch:2011ui},
while the second class flows from conformal $AdS_4$ to the same near-horizon Nernst brane geometry.
These flows do not satisfy the first-order BPS equations,
and hence are non-supersymmetric (they also do not obey the non-supersymmetric modification of the first-order flow equations mentioned above, but they do solve the second order equations of motion). They establish the path to resolve, within ${\cal N}=2$ gauged supergravity, 
the tidal force singularity at the Lifshitz center, which is 
now replaced by a Nernst brane near-horizon geometry. Admittedly, the horizon of the Nernst brane also has a tidal force singularity. 
However, this can be resolved at the classical level by heating up the brane. This transfers the singularity
to the Cauchy horizon cloaked by a smooth non-extremal event horizon \cite{Goldstein:2014qha,Dempster:2015xqa}. In this sense the singularity is of the ``good'' type in the Gubser classification \cite{Gubser:2000nd}. We leave this 
final step for future work. Moreover, the flows from the asymptotic Lifshitz geometry to the near-horizon Nernst geometry might lead
to a better understanding of the holographic embedding of this class of Nernst branes, which can now be 
seen as states in the Schr\"odinger group representation of the dual field theory.\footnote{Given that the Nernst brane does not carry any 
transverse momentum, we do not expect the effects discussed in \cite{Keeler:2013msa} to prevent a description of the 
Nernst brane using the boundary field theory.}

\section{First-order flow equations for static charged configurations}
\setcounter{equation}{0}
\label{sec:rewriting}

In this section, we derive first-order flow equations for static charged configurations in 
${\cal N}=2$ gauged supergravity theories in four dimensions with $n_V$ vector multiplets and $n_H$ hypermultiplets.
We denote the  $4 n_H$ real scalars  residing in the hypermultiplets by  $q^{\alpha} \, (\alpha = 1, \dots, 4 n_H)$.
These are
local coordinates of a quaternionic K\"ahler manifold, and the 
gaugings we consider are gaugings of the Abelian isometries of this manifold.
We denote the complex scalar fields residing in the vector multiplets by $z^A \, (A = 1, \dots, n_V)$.
However, rather than directly working with $z^A$, 
we work in the big moduli space with local complex coordinates $X^I \, (I = 0, \dots, n_V)$, 
in terms of which $z^A = X^A / X^0$.
The associated bosonic part of the Lagrangian is given in \eqref{action4d}. We refer to
appendix \ref{gaugsug} for details on ${\cal N}=2$ gauged supergravity Lagrangians.


\subsection{Electrically charged configurations}
\label{electric_rewrite}

In this subsection we will focus on static electrically charged configurations.
The configurations we seek are specified by the following space-time line element
\begin{equation}
ds^2 = - {\rm e}^{2 U} \, dt^2 + {\rm e}^{2 A} \left( dr^2 + dx^2 + dy^2 \right) \;,
\label{eq:line-blackb}
\end{equation}
with $U = U(r), \, A=A(r)$.  These configurations are supported by scalar fields $X^I (r), q^{\alpha}(r)$ 
as well as electric fields $E^I (r) = - \partial_r A^I_t (r)$.
Inserting this ansatz  into the four-dimensional Lagrangian $(-\sqrt{-g} L)$ given in \eqref{action4d} yields the 
following one-dimensional effective Lagrangian,
\begin{eqnarray}
\label{1d-effaction1}
{\cal L}_{1d}&=&  e^{\psi} \left[ U'^2 - \psi'^2 + A'^2 + e^{-2A} \, N_{IJ} Y'^I {\bar Y}'^J 
+  \Sigma_r^2 
+ \tfrac12 h_{\alpha {\beta}} \, q'^{\alpha} {q}'^{ \beta} 
\right. \\
&& \left. \qquad  - \tfrac12 e^{2(A-U)} \, 
h_{\alpha { \beta}} \,  (k^{\alpha}_I \, A^I_t)
(k^{{\beta}}_J \, A^J_t) 
+ \tfrac12 \, e^{-2U} \, {\rm Im} {\cal N}_{IJ} \, E^I E^J + e^{2A} V({X}, \bar{X}, q)  \right] + {\rm T.D.} \;, \nonumber
\end{eqnarray}
where $' = \partial_r$. The $Y^I$ denote rescaled scalar fields $X^I $, c.f. \eqref{AYY}, and $ \Sigma_r$ is given in \eqref{sigmaX}. We also introduced
\begin{equation}
\psi = A + U \;,
\end{equation}
for convenience. 
${\rm T.D.}$ denotes a total derivative. 
We refer to appendix \ref{apprew} for details on the derivation of \eqref{1d-effaction1}.

We would like to rewrite the one-dimensional effective Lagrangian \eqref{1d-effaction1} as a sum/difference
of squares that involve first-order derivatives.
In the absence of gauging, the rewriting would proceed by introducing electric charges and 
expressing the electric fields $E^I$ in terms of these
charges. In the presence of gauging, we follow this procedure, introducing a set of real fields $C_I$ and expressing
the electric fields $E^I$ in terms of them. In the absence of hypermultiplets, the $C_I$ reduce to electric charges dressed by warp
factors \cite{Barisch:2011ui}.
Furthermore, since the scalar fields $Y^I$ are complex, the rewriting 
will be achieved by means of complex combinations that we denote by $q_I$,\footnote{The quantity $q_I$ should not be confused with the hyper scalars 
$q^{\alpha}$.}
\begin{equation}
q_I = e^{i \gamma(r)} \left(C_I - i P^3_I \right) \;,
\label{def-qI}
\end{equation}
where $\gamma$ is an $r$-dependent phase and $P^3_I$ denotes one of the three real Killing prepotentials
appearing in the potential $V$ given in \eqref{pot}. Here we have opted to select $P^3_I$, but we could have
equally selected either $P^1_I$ or $P^2_I$. 
The significance of the phase $\gamma$ is as follows. In the context of ${\cal N}=2$ $U(1)$ gauged supergravity
(a gauged supergravity theory without hypermultiplets),
the combination $q_I Y^I$
can be viewed as describing an effective complex superpotential whose phase is $\gamma$ \cite{Dall'Agata:2010gj}.
The first-order flow equations are then derived from this superpotential. Hence, it is important to allow for the
presence of a phase $\gamma$, and this we have done in \eqref{def-qI}. Upon performing a first-order rewriting
of \eqref{1d-effaction1}, the phase $\gamma$ will turn out to satisfy a first-order flow equation, 
given in \eqref{Atil-BPS} below.

The rewriting  of the one-dimensional effective Lagrangian \eqref{1d-effaction1} as a sum/difference
of squares of first-order flow equations is described in detail in appendix \ref{apprew}.
The rewriting results in 
\begin{equation}
{\cal L}_{1d} = {\rm T.D.} + {\cal L}_{\rm squares} + \Delta \;,
\label{1d-effaction_rewrite}
\end{equation}
where
$\rm T.D.$ denotes a total derivative term, 
${\cal L}_{\rm squares}$ denotes a  sum/difference of squares, 
\begin{eqnarray}
{\cal L}_{\rm squares} &=& e^{\psi} \left[
\left(U' +  {\rm Re} \left( Y^I q_I \right) + e^{A-U} A_t^I C_I \right)^2 
 -  \left(\psi' + 2  {\rm Re} \left( Y^I q_I \right) + e^{A-U} A_t^I C_I \right)^2 \right.
 \nonumber\\
&&\quad  \left. + \left(A' +  {\rm Re} \left( Y^I q_I \right)\right)^2
  + e^{-2A} \, N_{IJ} \left( Y'^I  - e^{2A} N^{IK} {\bar q}_K \right) 
 \left( {\bar Y}'^J - e^{2A} N^{JL} q_L \right) 
+ \Sigma_r^2 \right. \nonumber\\
&&\quad  
 + \tfrac12 h_{{\alpha} {\beta}} \left( q'^{\alpha} - 2 \, h^{\alpha {\gamma}} \,
 {\rm Im} \left( e^{i \gamma} \, Y^I  \right)
\nabla_\gamma P^3_I 
  \right)
  \left( q'^{{\beta}} - 2 \, h^{\beta \delta }\,
 {\rm Im} \left( e^{i \gamma} \, Y^J  \right)
\nabla_\delta P^3_J
  \right)
  \nonumber\\
  && \quad \left. +  \tfrac12 \, e^{-2U} \, {\rm Im} {\cal N}_{IJ} \, \left( E^I + e^{U + A} ({\rm Im} {\cal N})^{IK} \, C_K \right)
   \left( E^J + e^{U + A} ({\rm Im} {\cal N})^{JL} \, C_L \right) 
   \right] \;, \nonumber\\
\end{eqnarray}
while $\Delta$ is given by 
\begin{eqnarray}
\Delta &=& e^{\psi} \left[ 2 \Lambda^- \left( \gamma' +  \Lambda^+ \right) - \tfrac12 
e^{2(A-U)}  \, h_{\alpha {\beta}} \,
(k^{\alpha}_I {\tilde A}_t^I)  \, (k^{ \beta}_J {\tilde A}_t^J) - e^{A - U} \, C_I' \, {\tilde A}^I_t 
 \right. \nonumber \\ 
&& \qquad  + 2 e^{A-U} \left(  h_{\alpha {\beta}} \, ( k_I^{\alpha}  {\tilde A}_t^I ) \,  k_J^{ \beta}   \, {\rm Re} (Y^J e^{i \gamma}) 
+ C_I {\tilde A}_t^I \, {\rm Re} (q_J Y^J)
\right) \nonumber\\
   && \qquad  
   \left. +  \left(e^{2A} \,  N^{IJ} - 2 Y^I {\bar Y}^J \right) \left(P^1_I P^1_J + P^2_I P^2_J \right) 
  \right]
 \;,
 \label{Delta_final}
\end{eqnarray}
where we introduced the combinations
\begin{eqnarray}
{\tilde A}_t^I &=& A_t^I + 2 e^{U-A} \, {\rm Re} \left(Y^I e^{i \gamma} \right) \;, \nonumber\\
\Lambda^{\pm} &=& {\rm Im} \left( C_I Y^I e^{i \gamma} \right) \pm {\rm Re} \left(
P^3_I Y^I e^{i \gamma} \right) \;.
\label{comb-A}
\end{eqnarray}

Let us now consider the variation of \eqref{1d-effaction_rewrite} with respect to the various fields and
demand its vanishing. Demanding that the variation of ${\cal L}_{\rm squares}$ vanishes can be achieved by setting each
of the squares in  ${\cal L}_{\rm squares}$ to zero. This yields 
the first-order flow equations
\begin{eqnarray}
U' &=& - {\rm Re} \left( Y^I q_I \right) - e^{A-U} A_t^I C_I \;, \nonumber\\
\psi' &=& -  2  {\rm Re} \left( Y^I q_I \right) - e^{A-U} A_t^I C_I \;,
 \nonumber\\
 A' &=& -  {\rm Re} \left( Y^I q_I \right) \;, \nonumber\\
 Y'^I  &=& e^{2 A} N^{IK} {\bar q}_K \;, \nonumber\\
q'^{\alpha} &=&
2 \, h^{\alpha {\beta}} \,
 {\rm Im} \left( e^{i \gamma} \, Y^I  \right)
\nabla_\beta P^3_I \;,
\nonumber\\
E^I &=& -  e^{U + A} ({\rm Im} {\cal N})^{IK} \, C_K \;, \nonumber\\
\Sigma_r &=& 0 \;.
\label{squarzer}
 \end{eqnarray}
Using \eqref{AYY}, we infer the compatibility of the equations for $A', Y'^I$.

Next, we consider the variation of $\Delta$ with respect to the various fields. Since the Killing prepotentials
$P^1_I$ and $P^2_I$ enter quadratically, the variation of the last line of \eqref{Delta_final} can be made to vanish by 
setting
\begin{equation}
P^1_I = P^2_I = 0 \;.
\label{P1P2}
\end{equation}
This selects a submanifold of the quaternionic K\"ahler manifold. Note that on this submanifold $\nabla_\beta P^3_I = \partial_\beta P^3_I$.

We now demand the vanishing of the variations of the first two lines of $\Delta$ in \eqref{Delta_final} with respect to the various fields.
Varying with respect to $C_I$ yields equations that are satisfied if we set
\begin{eqnarray}
&& \Lambda^- = 0 \;, \nonumber\\
&&  \gamma' +  \Lambda^+ = 0 \;, \nonumber\\
&& {\tilde A}_t^I =0 \;.
\label{Atil-BPS}
\end{eqnarray}
Then, varying with respect to $A_t^I$ yields 
\begin{equation}
C_I' = 2
h_{\alpha {\beta}} \,  k_I^{\alpha}  \,  k_J^{\beta}   \, {\rm Re} (Y^J
e^{i \gamma}) 
+ 2 C_I
\, {\rm Re} \left( q_J Y^J \right)  \;.
\label{C-eq}
\end{equation}
And finally, varying with respect to $\gamma$ or any of the remaining fields, 
the resulting equations are satisfied provided we impose \eqref{Atil-BPS}
and \eqref{C-eq}.  

Note that \eqref{C-eq} can also be written as
\begin{equation}
\left(C_I \, e^{2A} \right)' = 2 e^{2A} \,
h_{\alpha {\beta}} \,  k_I^{\alpha}  \,  k_J^{\beta}   \, {\rm Re} (Y^J
e^{i \gamma}) \;,
\label{maxwelleq}
\end{equation}
and that 
the consistency of $E^I = - \partial_r A_t^I$ with \eqref{squarzer} imposes the following relation,
\begin{equation}
({\rm Im} {\cal N})^{IJ} C_J = - 2 e^{-2A} \left[ \left(U' - A'\right) \, {\rm Re} \left(Y^I e^{i \gamma} \right)
+ {\rm Re} \left( Y'^I e^{i \gamma} \right) - \gamma' \, {\rm Im} \left( Y^I e^{i \gamma} \right)
\right] \;.
\label{eq2_C}
\end{equation}
This has to be consistent with \eqref{C-eq}.

Next, let us analyze the condition $\Sigma_r =0$ in \eqref{squarzer}.
Using \eqref{sigmaX} and the flow equations for $Y'^I$ and for $A'$ in \eqref{squarzer}, we compute
\begin{equation}
\Sigma_r = - i \, {\rm Im} \left( Y^I q_I \right) = - i \, \Lambda^- \;.
\end{equation}
Thus, the condition $\Sigma_r=0$ is equivalent to $\Lambda^- = 0$. We proceed to analyze the latter.
We will now show that the condition $\Lambda^- =0$ implies that the flow equation for $\gamma$ is automatically satisfied,
provided that ${\rm Re} (q_I Y^I) \neq 0$. Namely, starting from $\Lambda^- = {\rm Im }  ( q_I Y^I) = 0$ and differentiating with 
respect to $r$ results in
\begin{eqnarray}
\left[{\rm Im }  ( q_I Y^I)\right]'  = {\rm Im} \left[ i \gamma'\, q_I Y^I + e^{i \gamma} \left( C'_I Y^I -i (P^3_I)' \, Y^I\right) + q_I Y'^I \right] = 0 \;.
\end{eqnarray}
The term $q_I Y'^I$ is real by virtue of the flow equation for $Y'^I$. Using \eqref{C-eq}, we obtain
\begin{eqnarray}
\left[{\rm Im }  ( q_I Y^I)\right]'  &=& 
\left[ \gamma' + 2 {\rm Im} \left(C_I Y^I \, e^{i \gamma} \right) \right] \, {\rm Re} \left(q_J Y^J \right) \nonumber\\
&&- (P^3_I)' \, {\rm Re} \left(Y^I  e^{i \gamma}  \right) + 2 h_{\alpha \beta} \,k^{\alpha}_I \,k_J^{\beta} \,
 {\rm Im} \left( Y^I 
e^{i \gamma} \right)
{\rm Re} \left( Y^J e^{i \gamma} \right) \;.
\label{gammlam}
\end{eqnarray}
Using $(P^3_I)' = q'^{\alpha}  \nabla_{\alpha} P^3_I$, the flow equation for $q^{\alpha}$ and the relation 
\eqref{relPk}, we obtain that the second line of \eqref{gammlam} vanishes. It follows that 
\begin{eqnarray}
 \gamma' + 2 {\rm Im} \left(C_I Y^I \, e^{i \gamma} \right) =0 \,,
 \label{eq:flow-gam-alt}
 \end{eqnarray}
 provided  ${\rm Re} (q_I Y^I) \neq 0$.  Using $\Lambda^- =0$, we see that equation \eqref{eq:flow-gam-alt} is 
 precisely the flow equation for $\gamma$ derived above.

Using
the expression for $A_t^I$ in \eqref{comb-A}, we note that
the flow equations for $U$ and $\psi$ in \eqref{squarzer} can also be written as 
\begin{eqnarray}
U' &=&   {\rm Re} \left( Y^I C_I \, e^{i \gamma} \right) - 
{\rm Im} \left( Y^I \, P^3_I \, e^{i \gamma} \right) 
 \;, \nonumber\\
\psi' &=& -  2 \,  {\rm Im} \left( Y^I P^3_I \, e^{i \gamma} \right)\;.
\label{flowupsi}
\end{eqnarray}

This concludes the discussion of the first-order flow equations derived from \eqref{1d-effaction_rewrite}.
However, 
in addition to analyzing the conditions arising from the one-dimensional effective Lagrangian \eqref{1d-effaction_rewrite}, one has to make sure that also the Hamiltonian constraint is fulfilled.
By using the line element \eqref{eq:line-blackb},  
the Hamiltonian constraint turns out to coincide with the equation that is obtained by varying the effective Lagrangian
with respect to 
$U$. Thus, the Hamiltonian constraint is satisfied.

The flow equations \eqref{squarzer}, the Maxwell equation \eqref{maxwelleq}
and conditions \eqref{Atil-BPS} were first derived in the BPS context in \cite{Halmagyi:2013sla}. 
They can be generalized to first-order flow equations for non-supersymmetric configurations
in the following way.
Inspection of 
\eqref{E-BPS} shows that if one rotates $(E^I, C_I)$ by a constant invertible matrix $M$ according to
$(E^I, C_I) \rightarrow ((M E)^I, ((M^{-1})^T C)_I)$ such that $M^T {\rm Im} {\cal N} M =  {\rm Im} {\cal N}$,
one obtains a similar rewriting of the one-dimensional effective Lagrangian, but now  based on $q_I = e^{i \gamma} \left(((M^{-1})^T C)_I - i P^3_I \right)$. This allows one to obtain non-supersymmetric configurations from supersymmetric ones, as first observed
in \cite{Ceresole:2007wx}. We will return to this observation in section
\ref{sec:interpolating} when discussing interpolating solutions.
\label{sec:nonsusy}


\subsection{Adding magnetic charge density}

Now we add magnetic charges to the configuration specified by the line element \eqref{eq:line-blackb}.
This configurations will thus be supported by
flowing 
scalar fields $Y^I(r), q^{\alpha}(r)$
and by electric fields $E^I(r) = F^I_{tr} = - \partial_r A^I_t$ as well as magnetic fields
$F^I_{xy} = p^I$, where the $p^I$ are constant parameters that correspond to magnetic charge densities.  The associated non-vanishing gauge potentials are
\begin{equation}
(A^I_t, A^I_x = - \tfrac12 p^I y, A^I_y =  \tfrac12 p^I x) \;.
\end{equation}
Using \eqref{eq:line-blackb}, we then obtain for \eqref{kin-quat},
\begin{eqnarray}
h_{\alpha \beta} \, D_{\mu} q^{\alpha} \, D^{\mu} q^{\beta} &=& e^{-2A} \, h_{\alpha \beta} 
 \, q'^{\alpha} \, q'^{\beta} 
+ \, e^{-2U}
h_{\alpha \beta} \,  (k^{\alpha}_I \, A^I_t)
(k^{\beta}_J \, A^J_t) \nonumber\\
&& + \tfrac14 ( y^2  + x^2 ) \, e^{-2A} \, h_{\alpha \beta} \, (k^{\alpha}_I \, p^I) (k^{\beta}_J \, p^J)
\;,
\label{qut-metric-xy}
\end{eqnarray}
where $q'^{\alpha} = \partial q^{\alpha}/\partial r$. Observe that in the presence of the magnetic charge densities $p^I$,
there are $(x,y)$-dependent terms in \eqref{qut-metric-xy}, which can be removed by imposing 
\begin{equation}
p^I \, k_I^{\alpha} = 0 \;.
\label{constr-p}
\end{equation}
We will thus impose this constraint in the following.

Let us denote the combination $C_I e^{2A}$ appearing on the left hand side of \eqref{maxwelleq} by $Q_I$.
It is well known \cite{Ferrara:1997tw} that in the presence of magnetic charge densities, 
the term $- \frac12 Q_I ({\rm Im} {\cal N})^{IJ} Q_J$ in 
the Maxwell Lagrangian \eqref{E-BPS} gets extended to 
\begin{equation}
V_{\rm BH} = g^{A \bar{B}} {\cal D}_A Z \bar{\cal D}_{\bar B} {\bar Z} + |Z|^2 = 
\left( N^{IJ} + 2 {X}^I \bar{{X}}^J \right) {\hat Q}_I  \, \bar{\hat Q}_J \;,
\end{equation}
where
\begin{equation}
Z = - \hat{Q}_I \, X^I \;\;\;,\;\;\; \hat{Q}_I = Q_I - F_{IJ} \, p^J \;.
\end{equation}
Hence, we may readily incorporate magnetic charge densities by replacing
$C_I$ in \eqref{def-qI} with
\begin{equation} \label{Chat}
\hat{C}_I = C_I - F_{IJ}  \, p^J \, e^{-2A} \;.
\end{equation}
The rewriting of the one-dimensional effective action proceeds as before, but now in terms of 
\begin{equation}
q_I = e^{i \gamma} \left(\hat{C}_I - i P^3_I \right) \;.
\label{qChat}
\end{equation}
The electric field $E^I$ becomes expressed in terms of $C_I$ and $p^I$ as
\begin{equation} \label{EI}
E^I =- e^{U + A} ({\rm Im} {\cal N})^{IJ} \, \left( C_J   + {\rm Re} {\cal N}_{JK} \, p^K \, e^{-2A} \right)  \; ,
\end{equation}
cf.\ appendix \ \ref{sec:details_magnetic_charges} for a few more details on the derivation. 
The quantity $\Delta$ acquires an additional term, namely $- e^{\psi} \, P^3_I \, p^I$. Varying this term with respect
to $\psi$ yields the constraint
\begin{equation}
 P^3_I \, p^I =0 \;.
 \end{equation}
The quantities ${\tilde A}_t^I$ and $\Lambda^{\pm}$ are given by \eqref{comb-A}, with $C_I$ replaced by ${\hat C}_I$.

The resulting flow equations and constraints are summarized as follows,
\begin{eqnarray}
U' &=&   {\rm Re} \left( Y^I \hat{C}_I \, e^{i \gamma} \right) - 
{\rm Im} \left( Y^I \, P^3_I \, e^{i \gamma} \right) 
 \;, \nonumber\\
\psi' &=& -  2 \,  {\rm Im} \left( Y^I P^3_I \, e^{i \gamma} \right)\;, \nonumber\\
 A' &=& -  {\rm Re} \left( Y^I q_I \right) \;, \nonumber\\
 Y'^I  &=& e^{2 A} N^{IK} {\bar q}_K \;, \nonumber\\
q'^{\alpha} &=&
2 \, h^{\alpha {\beta}} \,
 {\rm Im} \left( e^{i \gamma} \, Y^I  \right)
\nabla_\beta P^3_I \;,
\nonumber\\
E^I &=& - e^{U + A} ({\rm Im} {\cal N})^{IJ} \, \left( C_J   + {\rm Re} {\cal N}_{JK} \, p^K \, e^{-2A} \right)  \;, \nonumber\\
\left(C_I \, e^{2A} \right)' &=& 2 e^{2A} \,
h_{\alpha {\beta}} \,  k_I^{\alpha}   \, k_J^{\beta} \, {\rm Re} (Y^J
e^{i \gamma}) \;, \nonumber\\
{A}_t^I &=& - 2 e^{U-A} \, {\rm Re} \left(Y^I e^{i \gamma} \right) \;, 
\nonumber\\
\gamma' &=&-  \Lambda^+  \;, \nonumber\\
\Sigma_r &=& \Lambda^- = 0 \;, \nonumber\\
p^I \, k_I^{\alpha} &=& 0 \;, \nonumber\\
p^I \, P^3_I  &=&0 \;, \nonumber\\
P_I^1 &=& P_I^2\,  = 0 \; ,
\label{eomcond}
 \end{eqnarray}
 with $q_I$ given by \eqref{qChat}, and with $\Lambda^{\pm}$ given by \eqref{comb-A}, with $C_I$ replaced by 
 ${\hat C}_I$.
As before, the conditions $\Sigma_r =0$ and $\Lambda^-=0$ are equivalent, and the latter implies the flow
equation for $\gamma$ provided ${\rm Re} (q_I Y^I) \neq 0$.
 The consistency of $E^I = - \partial_r A_t^I$ now imposes the following relation,
\begin{equation}
({\rm Im} {\cal N})^{IJ}  \left( C_J   + {\rm Re} {\cal N}_{JK} \, p^K\, e^{-2A} \right)  = 
- 2 e^{-2A} \left[ \left(U' - A'\right) \, {\rm Re} \left(Y^I e^{i \gamma} \right)
+ {\rm Re} \left( Y'^I e^{i \gamma} \right) - \gamma' \, {\rm Im} \left( Y^I e^{i \gamma} \right)
\right] \;.
\label{max2}
\end{equation}
The above equations were first derived in \cite{Halmagyi:2013sla} in the BPS context.

\section{Backgrounds}
\setcounter{equation}{0}
\label{sec:backgrounds}

Next, we consider solutions to the equations \eqref{eomcond}. 
We focus on solutions that are supported by
constant scalars ${\tilde X}^I$ and $q^{\alpha}$. The scalar ${\tilde X}^I$ is related to $X^I$ by  \eqref{XtilX}.


\subsection{Lifshitz backgrounds}

The Lifshitz line element is specified by
\begin{equation}
e^U = r^{-z} \;\;,\;\;\; e^A = r^{-1} \; ,
\end{equation}
such that the center is at $r \rightarrow \infty$ and the asymptotic region is at $r \rightarrow 0$.
Below we will show that $z \geq 1$.

We consider backgrounds that are supported by electric fields ($p^I =0$).
Inspection of \eqref{comb-A} yields
\begin{equation}
A_t^I = - 2 r^{-z} \, \, {\rm Re} \left({\tilde X}^I e^{i \gamma} \right) \;.
\label{Atlif}
\end{equation}
Inserting these expressions into the flow equation for $U$ yields
\begin{equation}
z = {\rm Re} \left( q_I {\tilde X}^I \right) - 2 {\rm Re} \left( C_I {\tilde X}^I \, e^{i \gamma}\right) 
= -  {\rm Re} \left( C_I {\tilde X}^I \, e^{i \gamma}\right) + {\rm Im} \left( P^3_I {\tilde X}^I \, e^{i \gamma}\right) 
\;.
\end{equation}
The flow equation for $A$ yields
\begin{equation}
{\rm Re} \left( q_I {\tilde X}^I \right) = 1 \;,
\label{cond1}
\end{equation}
and hence
\begin{equation}
z = 1  - 2 {\rm Re} \left( C_I {\tilde X}^I \, e^{i \gamma}\right) \;.
\label{value-z}
\end{equation}
Since $z$ is constant, we also take $C_I$ and $\gamma$ to be constant, in which case $q_I$ is also constant.
This is consistent with the 
flow equation for $Y^I$,
\begin{equation}
{\tilde X}^I = - N^{IJ} \, \bar{q}_J \;.
\label{flowX}
\end{equation}
Taking the real part of ${\tilde X}^I e^{i \gamma}$ in \eqref{flowX} gives
\begin{equation}
{\rm Re} \left( {\tilde X}^I \, e^{i \gamma} \right) = - N^{IJ} \, C_J \;,
\label{relXC}
\end{equation}
and hence, using \eqref{Atlif}, we obtain
\begin{equation}
A_t^I = 2 r^{-z} \, \, N^{IJ} C_J \;.
\end{equation}

Since we take $\gamma$ to be constant, we have $\Lambda_+ =0$ which, together with $\Lambda^-=0$, results in
\begin{eqnarray}
{\rm Im} \left( {C}_I {\tilde X}^I e^{i \gamma} \right) &=& 0 \;, \nonumber\\
{\rm Re} \left( P^3_I {\tilde X}^I e^{i \gamma} \right)  &=& 0 \;.
\label{cond-lamb}
\end{eqnarray}
Inspection of \eqref{max2} then yields 
\begin{eqnarray}
({\rm Im} {\cal N})^{IJ}   C_J    &=&  2 \, z  \, {\rm Re} \left({\tilde X}^I e^{i \gamma} \right) \;,
\label{maxz}
\end{eqnarray}
Contracting with $P^3_I$ we obtain,
\begin{equation}
P^3_I ({\rm Im} {\cal N})^{IJ} C_J =  0
\label{w-c-cond}
\end{equation}
due to \eqref{cond-lamb}.

Using the equation for $(C_I e^{2A})'$ in \eqref{eomcond} we obtain the relation
    \begin{equation}
 C_I = -  h_{\alpha \beta} \, k_I^{\alpha} \, k_J^{\beta}\, {\rm Re} \left({\tilde X}^J e^{i \gamma}\right) \;.
 \label{valueC1}
  \end{equation}
Contracting \eqref{valueC1} with   $ {\tilde X}^I e^{i \gamma}$ results in 
\begin{eqnarray}
h_{\alpha \beta} \, k_I^{\alpha} \, k_J^{\beta} \, {\rm Im} \left( {\tilde X}^I e^{i \gamma} \right) {\rm Re} \left({\tilde X}^J e^{i \gamma}\right) &=& 0 \;, \nonumber\\
{\rm Re} \left(C_I {\tilde X}^I e^{i \gamma} \right) &=& - h_{\alpha \beta} \, k_I^{\alpha} \, k_J^{\beta} \,
 {\rm Re} \left( {\tilde X}^I e^{i \gamma}\right)
  {\rm Re} \left( {\tilde X}^J e^{i \gamma}\right) \;,
\end{eqnarray}
where we used \eqref{cond-lamb}. 
Combining with \eqref{value-z}, we obtain
\begin{equation}
z = 1 + 2 h_{\alpha \beta} \, \, k_I^{\alpha} \, k_J^{\beta} \,
 {\rm Re} \left( {\tilde X}^I e^{i \gamma}\right)
  {\rm Re} \left( {\tilde X}^J e^{i \gamma}\right) 
\geq 1\;.
\end{equation}

Contracting \eqref{flowX} with $C_I$ we obtain
\begin{equation}
C_I {\tilde X}^I \, e^{i \gamma} = - C_I N^{IJ} \, \left( C_J + i P^3_J \right)\;.
\end{equation}
Using \eqref{cond-lamb}, \eqref{w-c-cond} and \eqref{ImNN}, we obtain
\begin{eqnarray}
C_I N^{IJ} P^3_J &=& - \left(C_I {\tilde X}^I \, P^3_J {\bar{\tilde X}}^J + 
P^3_I {\tilde X}^I \, C_J {\bar{\tilde X}}^J 
\right) \nonumber\\
&=& - {\rm Re} \left( C_I {\tilde X}^I \, e^{i \gamma} \right) 
\left[ {\rm Im} \left( P^3_I {\tilde X}^I \, e^{i \gamma} \right) + 
 {\rm Im} \left( P^3_I \bar{\tilde X}^I \, e^{-i \gamma} \right) \right] =0 \;.
\end{eqnarray}
Hence
\begin{equation}
{\rm Re} \left( C_I {\tilde X}^I \, e^{i \gamma} \right) = - C_I N^{IJ} C_J \;.
\end{equation}
Similarly, we infer
\begin{equation}
{\rm Im} \left( P^3_I {\tilde X}^I \, e^{i \gamma} \right) = - P^3_I N^{IJ} P^3_J \;.
\end{equation}
Thus, we may write \eqref{cond1} as
\begin{equation}
{\rm Re} \left( C_I {\tilde X}^I \, e^{i \gamma} \right) + 
{\rm Im} \left( P^3_I {\tilde X}^I \, e^{i \gamma} \right) =
- C_I N^{IJ} C_J - P^3_I N^{IJ} P^3_J =1 \;.
\end{equation}
And finally, combining \eqref{relXC} and \eqref{maxz} gives
\begin{equation}
(z-1) \, N^{IJ} C_J = \tilde{X}^I \, \left( C_J \bar{\tilde{X}}^J \right) + 
\bar{\tilde{X}}^I \, \left( C_J {\tilde{X}}^J \right) \;.
\label{maxw2}
\end{equation}
The constancy of $q^{\alpha}$ requires
\begin{equation}
{\rm Im} \left( {\tilde X}^I \, e^{i \gamma}  \right)
 \frac{\partial P^3_I}{\partial {q}^{{\beta}}} =0 \:.
 \label{constq}
 \end{equation}

Summarizing, the Lifshitz backgrounds we constructed are described by the following conditions, 
\begin{eqnarray}
{\tilde X}^I &=& {\rm const} \;\;\;,\;\;\;  q^{\alpha} = {\rm const} \;\;\;,\;\;\;
C_I = {\rm const} \;\;\;,\;\;\; p^I = 0 \;\;\;,\;\;\; \gamma = {\rm const}
\;, \nonumber\\
e^U &=& r^{-z} \;\;,\;\;\; e^A = r^{-1} \;, \nonumber\\
{\tilde X}^I &=& - N^{IJ} \, {\bar q}_J \;, \nonumber\\
C_I &=& -  h_{\alpha \beta} \, k_I^{\alpha} \,  k_J^{\beta} \, {\rm Re} \left( 
{\tilde X}^J e^{i \gamma}\right) \;,
\nonumber\\
 A_t^I &=&  2 \, r^{-z} \, \, N^{IJ} C_J \;, \nonumber\\
&z =& 1 - 2\, {\rm Re} \left( C_I {\tilde X}^I \, e^{i \gamma}\right) 
= 1 + 2 h_{\alpha \beta} \,k_I^{\alpha} \,k_J^{\beta} \,
 {\rm Re} \left( {\tilde X}^I e^{i \gamma}\right)
  {\rm Re} \left({\tilde X}^J e^{i \gamma}\right)  
 \geq 1\;, \nonumber\\
(z-1) \, N^{IJ} C_J &=& \tilde{X}^I \, \left( C_J \bar{\tilde{X}}^J \right) + 
\bar{\tilde{X}}^I \, \left( C_J {\tilde{X}}^J \right)  \;, \nonumber\\
0 &=& h_{\alpha \beta} \, k_I^{\alpha} \, k_J^{\beta} \, {\rm Im} \left( {\tilde X}^I e^{i \gamma} \right) {\rm Re} \left({\tilde X}^J e^{i \gamma}\right) \;, \nonumber\\
C_I N^{IJ} P^3_J &=& 0 \;\;\;,\;\;\; 
{\rm Im} \left( {C}_I {\tilde X}^I e^{i \gamma} \right) = 0 \;\;\;,\;\;\;
{\rm Re} \left( P^3_I {\tilde X}^I e^{i \gamma} \right)  = 0 \;\;\;,\;\;\;
 {\rm Im} \left( {\tilde X}^I \, e^{i \gamma}  \right)
 \frac{\partial P^3_I}{\partial {q}^{{\alpha}}} =0 \:, \nonumber\\
{\rm Re} \left( C_I {\tilde X}^I \, e^{i \gamma} \right) &+& 
{\rm Im} \left( P^3_I {\tilde X}^I \, e^{i \gamma} \right) =
- C_I N^{IJ} C_J - P^3_I N^{IJ} P^3_J =1 \;.
\label{conditions-lif}
\end{eqnarray}
In the BPS case, conditions for Lifshitz backgrounds have been discussed before in \cite{Cassani:2011sv,Halmagyi:2011xh}.

Note that while there is a lower bound on the allowed value of $z$, namely $z \geq 1$, there does not appear to be an
upper bound. In the following, we turn to specific examples with $z=2$.


\subsubsection{Examples}

On the hypermultiplet side, we now restrict to the universal
hypermultiplet. The associated scalars parametrize the coset space $SU(2,1)/SU(2) \times U(1)$. This is
a K\"ahler manifold, and we employ two standard parametrizations of this manifold in terms of complex coordinates,
which we summarize in appendix \ref{quater-metric}.  We denote these complex coordinates by $q^i, \bar{q}^{\bar
\imath}, i=1,2$.
We focus on solutions that are supported by
constant scalars ${\tilde X}^I$ and $q^i$.

In the first example, we choose the parametrization \eqref{quat-metric}, and consider the gauging of the following two isometries \cite{Behrndt:2000ph,Halmagyi:2011xh},
\begin{eqnarray}
k_I =  i a_I \, \left( q^1 \frac{\partial}{\partial q^ 1} 
- {\bar q^{\bar 1}} \frac{\partial}{\partial {\bar q^{\bar 1}}} \right)
+ i b_I \, \left( q^2 \frac{\partial}{\partial q^2} 
- {\bar q^{\bar 2}} \frac{\partial}{\partial {\bar q^{\bar 2}}} \right) \;,
\label{gaug1}
\end{eqnarray}
whose associated Killing prepotentials are $P^x_I = a_I \, P^x_a + b_I \, P^x_b$,
with $P_a, P_b$ given by
 \begin{equation}
P_a =\frac{-2}{(|q^1|^2  + |q^2|^2) \sqrt{ 1- |q^1|^2-|q^2|^2 } }
 \begin{pmatrix}
- {\rm Im} (q^1 q^2)\\
{\rm Re} (q^1 q^2)\\
\frac{-|q^2|^4 + |q^2|^2 - |q^1|^2 (1 + |q^2|^2)}{2 \sqrt{1- |q^1|^2 - |q^2|^2}}
\end{pmatrix}
\end{equation} 
and
\begin{equation}
P_b =\frac{-2}{(|q^1|^2  + |q^2|^2) \sqrt{1- |q^1|^2-|q^2|^2 }} 
\begin{pmatrix}
{\rm Im} (q^1 q^2)\\
- {\rm Re} (q^1 q^2)\\
\frac{-|q^1|^4 + |q^1|^2 - |q^2|^2 (1 + |q^1|^2)}{2 \sqrt{1- |q^1|^2 - |q^2|^2}}\end{pmatrix} \;.
\end{equation} 
We keep $q^2$ fixed at $q^2 =0$, so that $P^1_I = P^2_I =0$. At $q^2=0$, $P^3_I$ is then given by
\begin{equation}
P^3_I = \frac{a_I}{1- |q^1|^2 } - b_I \;.
\label{kprep1}
\end{equation}

On the vector multiplet side, we consider 
the STU-model with prepotential $F(X) = - X^1 X^2 X^3 / X^0$ with $X^0$ real and $X^A$ imaginary ($A = 1,2,3$). 
For this case, the fields $S = - i X^1/X^0, T = - i X^2/X^0, U = -i X^3/X^0$ are all real, and
the matrix $N^{IJ}$ is given by
\begin{equation} \label{NIJmatrix}
N^{IJ}=  \frac{1}{4 STU}
 \begin{pmatrix}
1& 0 & 0 & 0  \\
0  &  S^2 & - S T & - S U  \\
0 & - S T & T^2 & - T U \\
0 & - S U & - T U  & U^2 
\end{pmatrix} \;.
\end{equation}

We set $\gamma =0$ and take non-vanishing charges and fluxes $(C_0; a_0, b_0, b_1, b_2, b_3)$.
{From} $C_I N^{IJ} P^3_J =0$ we obtain
\begin{equation}
C_0 N^{00} P^3_0 = 0 \rightarrow P^3_0 = \frac{a_0}{1 - |q^1|^2} - b_0 = 0 \;,
\label{valueq1}
\end{equation}
which implies $b_0 > a_0$ and expresses the value of $|q^1|$ in terms of fluxes $a_0, b_0$. 
Since $ k_I^i \, {\rm Im} \left( {\tilde X}^I e^{i \gamma}  \right) =  0$, the equation
$h_{\alpha \beta} \, k_I^{\alpha} \, k_J^{\beta} \, {\rm Im} \left( {\tilde X}^I e^{i \gamma} \right) {\rm Re} \left({\tilde X}^J e^{i \gamma}\right) =0$ is satisfied, and so are the equations
\begin{eqnarray}
{\rm Im} \left( {\tilde X}^I \, e^{i \gamma}  \right)
 \frac{\partial P^3_I}{\partial {q}^{{i}}} =0 \;\;\;,\;\;\;
{\rm Im} \left( {C}_I {\tilde X}^I e^{i \gamma} \right) = 0 \;\;\;,\;\;\;
{\rm Re} \left( P^3_I {\tilde X}^I e^{i \gamma} \right)  = 0 \;.
\end{eqnarray}
The equation 
\begin{equation}
{\tilde X}^0 = - N^{0J} \, {\bar q}_J = - N^{00} \, C_0 \;,
\label{eq:X0}
\end{equation}
when combined with $C_I = 
-  h_{1 \bar 1} \, |q^1|^2 \, a_I \, a_J \, {\rm Re} \left({\tilde X}^J e^{i \gamma}\right) $, results in
\begin{equation}
 h_{1 \bar 1} \, |q^1|^2 \, a_0^2 N^{00} = 1 \;,
\label{eq:valueq1}
\end{equation}
which relates the value of $|q^1|$ to the values of the scalar fields of the vector multiplets. 

The value of $z$ is given by
\begin{equation}
z = 1 + 2 C_0^2 \, N^{00} \;.
\label{zCN}
\end{equation}
This satisfies equation \eqref{maxw2}.

The equation \begin{equation}
{\tilde X}^A = - N^{AJ} \, {\bar q}_J = i N^{AB} \, b_B \;\;\;,\;\;\; A=1,2,3 \;.
\label{eq:XA}
\end{equation}
when combined with \eqref{eq:X0}, 
results in 
\begin{equation}
T^A = - i \frac{\tilde X^A}{\tilde X^0} = - \frac{N^{AB} b_B}{N^{00} C_0} \;.
\label{eq:valueT}
\end{equation}
This yields
\begin{equation}
S=\frac{C_0}{b_1} \;\;\;,\;\;\; 
T=\frac{C_0}{b_2} \;\;\;,\;\;\; 
U=\frac{C_0}{b_3} \;.
\label{valueSTU}
\end{equation}
Requiring $S, T, U > 0$ leads to
\begin{equation}
 \frac{ C_0 }{b_A} > 0 \;.
 \label{Cbrel}
 \end{equation}
The equation
\begin{eqnarray}
{\rm Re} \left( C_I {\tilde X}^I \, e^{i \gamma} \right) + 
{\rm Im} \left( P^3_I {\tilde X}^I \, e^{i \gamma} \right) =
- C_I N^{IJ} C_J - P^3_I N^{IJ} P^3_J =1 
\end{eqnarray}
gives
\begin{equation}
C_0^2 \, N^{00} = - 1 - b_A N^{AB} b_B \;.
\label{eq:relCBb}
\end{equation}
Inserting \eqref{valueSTU} into \eqref{eq:relCBb} gives
\begin{equation}
C_0 = \frac{b_1 b_2 b_3}{2}  \;,
\end{equation}
so that
\begin{equation}
C_0^2 \, N^{00} = \frac12 \;,
\end{equation}
and hence\footnote{We have checked that $z=2$ also holds when all the fluxes $(a_I, b_I)$ are turned on.}
\begin{equation}
z = 2\;.
\label{z2lif}
\end{equation}
Using the metric \eqref{quat-metric}, the relations \eqref{valueq1} and \eqref{eq:valueq1} result in the relation
\begin{equation}
b_0 (b_0 - a_0) = 4 \frac{C^3_0}{b_1 b_2 b_3} \;,
\end{equation}
which is consistent with the requirements $b_0 > a_0$ and \eqref{Cbrel}.
Thus, we have constructed a new $z=2$ Lifshitz background.

Next, we consider a different example, based on the parametrization \eqref{met-pet}
(with  $q^i=S, \xi$), and we gauge the following two isometries \cite{Halmagyi:2011xh},
\begin{equation}
k_I = a_I \, i \left(\frac{\partial}{\partial S} - \frac{\partial}{\partial \bar S} \right) + b_I \, i \, \left( \xi \frac{\partial}{\partial \xi} - {\bar \xi} \frac{\partial}{\partial \bar \xi} \right) \;.
\end{equation}
The associated Killing prepotentials are $P_I^x = a_I \, P_a^x + b_I \, P_b^x $ with
 \begin{equation}
P_a^x =
 \begin{pmatrix}
0 \\
0 \\
- \frac{1}{2 \rho}
\end{pmatrix}
\;\;\;,\;\;\;
P_b^x =
\begin{pmatrix}
\frac{ \xi + \bar \xi}{\sqrt{\rho}}\\
- i \frac{ \xi -  \bar \xi}{\sqrt{\rho}}\\ 
\frac{|\xi|^2}{\rho} - 1 
\end{pmatrix} \;.
\end{equation} 
We keep $q^2 = \xi$ fixed at $q^2 =0$, so that $P^1_I = P^2_I =0$. At $q^2=0$, $P^3_I$ is then given by 
\begin{equation}
P^3_I = - \frac{a_I}{2 \rho } - b_I \;.
\end{equation}

On the vector multiplet side, we consider 
the $t^3$-model with prepotential $F(X) = - (X^1)^3 / X^0$.
We set  $t = X^1 / X^0 = t_1 + i t_2$. The matrix $N^{IJ}$ is given by
\begin{equation}
N^{IJ}=  \frac{1}{4 t_2^3}
 \begin{pmatrix}
1& t_1 \\
t_1  &  t_1^2 -  \tfrac13 t_2^2 \\
\end{pmatrix} \;.
\end{equation} 
We set $\gamma$ to a constant, but arbitrary, value, and we take non-vanishing charges and fluxes $(C_0; a_0, b_1)$.
The equation ${\tilde X}^I = - N^{IJ} {\bar q}_J$ results in 
\begin{eqnarray}
e^{i \gamma } \, {\tilde X}^0 &=& - \frac{1}{4 t_2^3} \left( C_0 - i \frac{a_0}{2 \rho} - i t_1 b_1 \right) \;, \nonumber\\
e^{i \gamma } \, {\tilde X}^1 &=& - \frac{1}{4 t_2^3} \left( t_1 \left( C_0 - i \frac{a_0}{2 \rho}\right)  - i \left( t_1^2 - \tfrac13 t_2^2 \right)  b_1 \right) \;.
\label{valueX01}
\end{eqnarray}
Imposing ${\rm Im} \left( C_I {\tilde X}^I \, e^{i \gamma} \right) = 0$, we obtain
\begin{equation}
\frac{a_0}{2 \rho} + t_1 b_1 =0 \;,
\label{t1val}
\end{equation}
which determines the value of $\rho t_1$ in terms of the fluxes. Thus, $e^{i \gamma } \, {\tilde X}^0$ is real.
Using \eqref{t1val}, we determine from \eqref{valueX01} 
\begin{eqnarray}
t = \frac{t_1 \left( C_0 - i \frac{a_0}{2 \rho}\right)  - i \left( t_1^2 -\tfrac13 t_2^2 \right)  b_1  }{C_0} = t_1 + i  \frac{   t_2^2  \, b_1  }{3 C_0} \;,
\end{eqnarray}
from which we conclude
\begin{equation}
t_2 =  \frac{3 C_0}{b_1} \;,
\label{valuet2}
\end{equation}
which we demand to be positive, and hence
\begin{equation}
\frac{C_0}{b_1} > 0 \;.
\end{equation}
The condition ${\rm Re} \left( P^3_I {\tilde X}^I \, e^{i \gamma} \right) = 0$ is satisfied by virtue of \eqref{t1val},
and so are the conditions
$C_I N^{I J } P^3_J = 0 $, $ k_I^\alpha \, {\rm Im}  \left( {\tilde X}^I e^{i \gamma} \right) = 0 $ and 
${\rm Im} \left( {\tilde X}^I \, e^{i \gamma}  \right)
 \partial P^3_I/\partial {q}^{i} =0$.
The condition 
${\rm Re} \left( C_I {\tilde X}^I \, e^{i \gamma} \right) + 
{\rm Im} \left( P^3_I {\tilde X}^I \, e^{i \gamma} \right) =1$ gives
\begin{equation}
t_2 ^3 = \frac12 C_0^2 \;,
\label{valuet22}
\end{equation}
where we used \eqref{valuet2}. 

The equation for $C_0$ results in
\begin{equation}
C_0 = - h_{S \bar S} \, a_0 ^2  \, {\rm Re}\left(  {\tilde X}^0 e^{i \gamma } \right) \;.
\end{equation}
Using $h_{S \bar S} = 1/(4 \rho^2)$ this results in 
\begin{equation}
\rho^2 = \frac{a_0^2}{16 t_2^3} \:,
\end{equation}
which combined with \eqref{valuet22} gives 
\begin{equation}
\rho^2  =  \frac{a_0^2}{8 C_0^2}  \;.
\end{equation}
Thus, the only scalar whose value remains unfixed is the hyper scalar $S - \bar S$.

Finally, we compute the value of $z$, 
\begin{equation}
z = 1 - 2  \, {\rm Re} \left( C_I {\tilde X}^I  e^{i \gamma } \right) = 1 + \frac{C_0^2}{2 t_2^3} =  1 + 1 = 2 \,,
\end{equation}
and we check that the condition \eqref{maxw2} is satisfied.

This reproduces the BPS background constructed in \cite{Cassani:2011sv}.


\subsection{AdS backgrounds}

A possible way for resolving the tidal force singularity consists in searching for solutions that flow from $AdS_4$ to 
$AdS_2 \times \mathbb{R}^2$ passing through an 
intermediate Lifshitz region. Numerical flows of this kind have, for instance, been studied in the context of Einstein-Maxwell-dilaton systems
with loop or Weyl-squared corrections 
in \cite{Bhattacharya:2012zu,Knodel:2013fua}. Thus, it is of interest to investigate the construction of AdS backgrounds in the models at hand.
To this end, we return to the gauging \eqref{gaug1} with Killing prepotential \eqref{kprep1}. On the vector multiplet
side, we again consider the STU-model with prepotential  $F(X) = - X^1 X^2 X^3/X^0$. 

We first consider $AdS_4$ backgrounds. These are backgrounds satisfying \eqref{conditions-lif} with
$z=1$ and $C_I=0$, and supported by arbitrary fluxes $(a_I, b_I)$. The condition ${\tilde X}^I = - N^{IJ} {\bar q}_J$
then gives ${\rm Re} \left( {\tilde X}^I e^{i \gamma} \right) =0$. This, in turn, implies that the physical moduli
$S, T, U$ are all purely imaginary, which does not correspond to a sensible solution in supergravity.
Thus, we cannot construct $AdS_4$ backgrounds in this model.\footnote{This is similar to the findings of \cite{Halmagyi:2011xh} where also no $AdS_4$ solutions were found with only  electric gaugings. Note, however, that there are $AdS_4$ solutions with purely electric gaugings in a duality frame with a 
square root prepotential, cf.\ \cite{Hristov:2009uj,Halmagyi:2013sla}.}

Next, let us consider $AdS_2 \times \mathbb{R}^2$ backgrounds. These are solutions to the equations \eqref{eomcond} 
with constant 
scalar fields $Y^I$ and $q^{\alpha}$.
The associated line
element is specified by
\begin{equation}
U = r \;\;\;,\;\;\; A = {\rm constant} \;.
\end{equation}
Setting $Y'^I=0$ gives $q_I =0$, which results in ${\hat C}_I = i P^3_I$,
and hence
\begin{eqnarray}
C_I \, e^{2A} &=& \tfrac12 \left(F_{IJ} + {\bar F}_{IJ} \right)  p^J \;, \nonumber\\
P^3_I \, e^{2A} &=& - \tfrac12 N_{IJ} p^J \;.
\label{attr-ads}
\end{eqnarray}
The value of $e^{2A}$ is given by
\begin{equation}
e^{2A} = - N_{IJ} \, Y^I {\bar Y}^J \;.
\end{equation}
The $C_I$ are constant.
Since $q_I =0$, the condition $\Lambda^- = {\rm Im} (q_I Y^I) = 0$ is automatically satisfied. Demanding $\gamma'=0$,
the condition $\Lambda^+ =0$ yields
\begin{equation}
{\rm Re}  \left( e^{i \gamma} \, Y^I \, P^3_I \right) = 0 \;.
\label{reYP}
\end{equation}
The flow equation for $\psi$ yields
\begin{equation}
{\rm Im} \left( e^{i \gamma} \, Y^I \, P^3_I \right) = - \tfrac12 \;.
\label{eq-flowU}
\end{equation}
The constancy of $q^{\alpha}$ implies \eqref{constq}.

For the model at hand, we take the complex hyper scalars $q^i$ to have values $q^2 =0, q^1 \neq 0$.
We seek axion free solutions, i.e. solutions for which 
$Y^0$ is real and $Y^A$ are imaginary ($A=1,2,3$), so that $S = -i Y^1/Y^0, T = - i Y^2/Y^0, U = -i Y^3/Y^0$ are real.
We choose non-vanishing charges $(C_0, p^1, p^2, p^3)$.  Then, the attractor equations \eqref{attr-ads}
 yield
\begin{eqnarray}
C_0 \, e^{2A} &=& - \left(TU p^1 + SU p^2 + ST p^3  \right) \;, \nonumber\\
P^3_A \, e^{2A} &=& - \tfrac12 N_{AB} \, p^B \;\;\;,\;\;\; A,B =1,2,3 \;, \nonumber\\
P^3_0 &=& 0 \;.
\label{stuAC}
\end{eqnarray}
Taking $q^1\neq 0$, 
the condition $p^I k_I^{\alpha} = 0$ yields $p^I \,  a_I = 0$.
The condition $p^I P^3_I=0$ then gives $p^I b_I = 0$.
The condition \eqref{constq} results in 
\begin{equation}
a_I \, {\rm Im} \left( e^{i \gamma} \, Y^I \right) = 0 \;.
\label{aImY}
\end{equation}
Requiring $C'_I =0$ gives 
\begin{equation}
a_I \, {\rm Re} \left( e^{i \gamma} \, Y^I \right) = 0 \;,
\label{aY}
\end{equation}
cf.\ \eqref{C-eq} with $q_J=0$.
We take $\gamma =0$ in the following. Then, since the $Y^A$ are imaginary, the 
condition \eqref{aY} is satisfied provided $a_0=0$. It follows from $P^3_0 = 0$ that then also $b_0=0$.
Then, the conditions $p^I a_I = p^I b_I =0$ and equations \eqref{reYP}, \eqref{eq-flowU} and \eqref{max2} result in 
\begin{eqnarray}
p^1 a_1 + p^2 a_2 + p^3 a_3 &=& 0 \;, \nonumber\\
p^1 b_1 + p^2 b_2 + p^3 b_3 &=& 0 \;, \nonumber\\
{\rm Re}  \left( Y^A \, b_A \right) &=& 0 \;, \nonumber\\
{\rm Im}  \left( Y^A \, b_A \right) &=&  \tfrac12 \;, \nonumber\\
({\rm Im} {\cal N})^{IJ}  \left( C_J \, e^{2A}   + {\rm Re} {\cal N}_{JK} \, p^K \right)  &=& - 2  \, {\rm Re} \left(Y^I e^{i \gamma} \right) \;.
\end{eqnarray}
These equations, combined with \eqref{stuAC} and \eqref{aImY}, can be consistently solved.
For instance, taking $ a_2 = a_3 = -1,
b_2 = -1,   b_3 = 1, p^3 = 0.111$ and setting $Y^0=1$, we obtain 
\begin{eqnarray}
a_1 \approx 0.996\, b_1 \;\;,\;\;
p^1 \approx -56.270 / b_1 \;\;,\;\;
p^2 \approx -56.159
\end{eqnarray}
and
\begin{eqnarray}
|q^1| \approx 0.044 \;\;,\;\;
S \approx 42.328 / b_1 \;\;,\;\;
T \approx 41.995 \;\;,\;\; 
U \approx 0.167 \;\;,\;\; C_0 = \frac14 \; ,
\end{eqnarray}
where $b_1$ remains unfixed. Hence, we have constructed a new background.

Thus, we see that in the model at hand, when $\gamma =0$, it is not possible to construct an $AdS_2 \times \mathbb{R}^2$ background
carrying the same fluxes as the Lifshitz backgrounds constructed above which necessarily have $a_0 \neq 0$, cf.\ \eqref{eq:valueq1}.  This does not exclude the existence of 
both types of backgrounds but with different values of $\gamma$. An interpolating solution between two such vacua would
then require a varying $\gamma (r)$.\footnote{A non-supersymmetric solution in 5d gauged supergravity interpolating between asymptotic $AdS_5$ and $Lif_3$ in the center was given in \cite{Burda:2014jca}.}  Rather than pursuing this question, we turn to a different mechanism for cloaking
the singularity at the center of the Lifshitz background.


\section{Interpolating between $z=2$ Lifshitz and Nernst geometries}
\setcounter{equation}{0}
\label{sec:interpolating}

Lifshitz backgrounds with $z=2$ can be obtained from five dimensions by dimensional Scherk-Schwarz reduction
\cite{Chemissany:2012du}.
Let us thus consider the $z=2$  Lifshitz background of the STU-model described above \eqref{z2lif} and lift it to five dimensions.
For convenience we set $b_1b_2b_3 =2$, so that $C_0 =1$.\footnote{The following analysis can, however, easily be generalized to other values of the fluxes $b_1, b_2, b_3$.} Then, the four-dimensional line element and the gauge potential 
1-form $A^0 = A_t^0 dt$
read
\begin{eqnarray}
ds^2_4 &=& - \frac{dt^2}{r^4} + \frac{dr^2 + dx^2 + dy^2}{r^2} \;, \nonumber\\
A_t^0 &=& \frac{1}{r^2} \;.
\end{eqnarray}

This lifts to a solution in five dimensions with line element \cite{Chemissany:2012du}
\begin{eqnarray}
ds^2_5 = ds^2_4  + (du - A^0)^2 = 
\frac{dr^2}{r^2} + \frac{\left(- 2 dt \, du + dx^2 + dy^2\right)}{r^2} + du^2 \;,
\end{eqnarray}
where $u$ denotes the coordinate of the $S^1$ along which the Scherk-Schwarz reduction is performed.
The five-dimensional vector scalars $X^A_5$ supporting the solution are constant, and their
values are given by  $X^A_5 = -i Y^A/Y^0$ given in \eqref{valueSTU}.
The Proca mass term for $A^0$ in four dimensions arises by Scherk-Schwarz reduction of one of the hyper scalars
in five dimension.  Furthermore, the solution is uncharged in five dimensions.

We now deform the line element in five dimensions as follows,
\begin{eqnarray} \label{5dmetric}
ds^2_5 &=& 
\frac{dr^2}{r^2} + \frac{\left(- 2 dt \, du + dx^2 + dy^2\right)}{r^2} + f^{-2}(r) \,  du^2 \;,
\end{eqnarray}
which we write as
\begin{eqnarray} \label{metric5d}
ds^2_5 = f \left(- \frac{f \, dt^2}{r^4} + \frac{dr^2 + dx^2 + dy^2}{f \, r^2} \right)
 + f^{-2} \left(du - \frac{f^2}{r^2} dt\right)^2 \:.
\end{eqnarray}
Then, dimensionally reducing along the $S^1$
parametrized by $u$, we obtain 
\begin{eqnarray} \label{deform-line}
ds^2_4 &=& - \frac{f \, dt^2}{r^4} + \frac{dr^2 + dx^2 + dy^2}{f \, r^2} \;, \\
A_t^0 &=& \frac{f^2}{r^2} \;,
\end{eqnarray}
with the four-dimensional vector scalars $z^A = Y^A/Y^0$ now given by 
\begin{equation} \label{deform-scalars}
z^A = z^A_u/f\ ,
\end{equation}
where $z^A_u$ denote
the values of the vector scalars in the undeformed case corresponding to \eqref{valueSTU}. The values of the four-dimensional hyper scalars $q^i$ 
are not deformed.  Then the four-dimensional equations of motion are solved provided that $f$
is given by the following two-parameter family of deformations,
\begin{equation} \label{f}
f^2(r) = \frac{r^2}{r^2 + \alpha \, r^4 + \beta} \;\;\;,\;\;\; \alpha, \beta \in \mathbb{R} \;.
\end{equation}
Note that this four-dimensional solution does not satisfy the first-order flow equations \eqref{eomcond}, and hence,
is not supersymmetric. The five-dimensional metric \eqref{5dmetric} with $f$ given in \eqref{f} 
was first written down in \cite{Chemissany:2011mb}.\footnote{Metrics with a similar structure also appear in \cite{Balasubramanian:2010uk,Donos:2010tu}, where, however, the corresponding 
function $f$ does not depend on $r$, but on the analog of our coordinate $u$ along which one performs the reduction.} Here we would like to 
investigate what the reduced four-dimensional metric \eqref{deform-line} implies for our goal to cloak the Lifshitz tidal 
force singularity within ${\cal N}=2$ gauged supergravity. 

To this end, let us consider the behaviour of the deformed solution \eqref{deform-line}.
First, consider the case $\alpha \neq 0, \beta =0$. At $r \rightarrow 0$ (i.e.\ asymptotically), we have $f \rightarrow 1$, and the solution
approaches the $z=2$ Lifshitz solution described above \eqref{z2lif}. On the other hand, as $r \rightarrow \infty$ (i.e.\ at the center), we have $f^2 \rightarrow
(\alpha \, r^2)^{-1}$, and hence $\alpha$ must be taken to be positive.  As $r \rightarrow \infty$, 
the line element and the vector scalars $z^A$ asymptote to
\begin{eqnarray}
ds^2_4 &=& - \frac{dt^2}{\sqrt{\alpha} \, r^5} +  \sqrt{\alpha} \, \frac{dr^2 + dx^2 + dy^2}{r} \;, \nonumber\\
z^A &=& z^A_u \, \sqrt{\alpha} \,  r \;.
\label{deform-asym}
\end{eqnarray}
By changing the radial coordinate, $r = c /\sqrt{\tilde r}$ with $c = 4/(\alpha)^{1/2}$, and by rescaling $dt, dx$ and $dy$,
the deformed line element
is seen to describe the near-horizon geometry of the Nernst brane constructed in \cite{Barisch:2011ui},
\begin{eqnarray}
ds^2_4 &=& - {\tilde r}^{5/2} \, dt^2  +  \frac{d{\tilde r}^2}{{\tilde r}^{5/2}}  + {\tilde r}^{1/2}
\left( dx^2 + dy^2 \right)\;, \nonumber\\
z^A &=&  4 z^A_u \, {\tilde r}^{-1/2} \;.
\label{deform-asym2}
\end{eqnarray}
Next, let us consider the case $\alpha =0, \beta \neq 0$. When $r \rightarrow \infty$, we have $f \rightarrow 1$,
and hence we recover the Lifshitz background described above \eqref{z2lif}. However, this is where the Lifshitz metric exhibits a tidal
force singularity, and hence the interpolating solution does not help in addressing this issue.
We will thus discard this case.

Finally, consider the case $\alpha \neq 0, \beta \neq 0$. As $r \rightarrow \infty$, the solution has
the Nernst behavior \eqref{deform-asym2}. When $r \rightarrow 0$, $f$ approaches $f^2 \rightarrow r^2/\beta$, and hence
$\beta$ has to be positive.  In this limit we then obtain
\begin{eqnarray}
ds^2_4 &=& r^{-3} \left( - \frac{dt^2}{\sqrt{\beta}} + \sqrt{\beta} \, \left( dr^2 + dx^2 + dy^2 \right) \right) \;, \nonumber\\
z^A &=& z^A_u \, \sqrt{\beta} \, r^{-1} \;.
\end{eqnarray}
This line element describes a conformal AdS geometry. Thus, we have an interpolating solution that interpolates
between a conformal AdS geometry and the near-horizon geometry of a Nernst brane.

To summarize, \eqref{deform-line} -- \eqref{f} constitute a two-parameter class of Nernst brane backgrounds (for $\alpha \neq 0$)
which do not obey the first-order flow equations  \eqref{eomcond}, and hence are non-supersymmetric. In fact, they also do not solve the non-supersymmetric flow equations 
based on a constant matrix $M$
mentioned at the end of section \ref{electric_rewrite}. We show this in appendix \ref{sec:no_solution_flow_eqs}. The near-horizon geometry for these non-supersymmetric Nernst branes is exactly the same as 
the one of the supersymmetric Nernst brane of \cite{Barisch:2011ui}, indicating that this might be a universal feature of Nernst backgrounds, 
at least of those arising from gauged supergravity. 
We note that also the $\eta$ geometries of \cite{Hartnoll:2012wm} can be embedded in gauged supergravity, cf.\ \cite{Donos:2012yi}. For $\eta >0$ 
they also have the feature that the entropy vanishes in the zero temperature limit, according to $s \sim T^\eta$. However, 
in contrast to our Nernst brane solutions which have vanishing Ricci and Kretschmann scalar at the horizon but suffer from a tidal force singularity,
$\eta$ geometries with $\eta >0$ have curvature singularities at the horizon.\footnote{For example, the near-horizon behavior of the $\eta=1$ geometry is 
schematically given by $ds^2 = -r^a dt^2 + r^{-a} dr^2 + r^{1/2} (dx^2 + dy^2)$ with $a=3/2$ instead of the $a=5/2$ behavior of the Nernst brane, cf.\ \eqref{deform-asym2}. It is easy to show that for this metric the Ricci and Kretschmann scalars diverge at $r=0$ for $a<2$ (i.e.\ for the $\eta=1$ geometry) whereas they vanish for $a>2$ (i.e.\ for the Nernst brane geometry).} Thus, the $\eta$ geometries for $\eta>0$ behave more like small black holes which also have a curvature singularity at the horizon.
In the non-supersymmetric case discussed here, the near-horizon Nernst background turns out to be the common IR 
fixed point of flows with two different types of UV behaviors corresponding to Lifshitz ($\beta =0$) 
and conformal AdS ($ \beta \neq 0$) backgrounds.\footnote{In five dimensions, the constant $\beta$ can be set to zero by the coordinate transformation 
$t \rightarrow t + \beta u/ 2 $, as was also noticed in \cite{Chemissany:2011mb}. However, this is not a viable coordinate transformation in the 
four-dimensional effective theory obtained by keeping only the Kaluza-Klein zero modes. Thus, different values for $\beta$ do lead to different physics in the 
four-dimensional effective theory.}

Both these backgrounds have divergent scalar fields (in addition to the tidal force singularity at the center), and this indicates that the correct
frame to view them is in five dimensions, as was also the case for the supersymmetric Nernst brane 
\cite{Barisch:2011ui} whose five-dimensional uplift was discussed in \cite{BarischDick:2012gj}.
When lifted up to five dimensions, the Lifshitz background becomes a Poincar\'e breaking solution, while
the conformal AdS background lifts up to $AdS_5$, cf.\ eq.\ \eqref{metric5d}. We expect that the scalar fields can be regularized by
heating up the system, as was the case for the non-extremal version of the supersymmetric Nernst brane derived in \cite{Dempster:2015xqa}.

\section{Conclusions}
\setcounter{equation}{0}
 Lifshitz geometries are plagued with a tidal force singularity at their center, which  needs to be dealt with,
for instance by constructing interpolating solutions that flow away from Lifshitz to other solutions. Hence, 
the primary goal of this paper was to examine the space of solutions interpolating between non-relativistic vacua such as $z=2$ Lifshitz and other solutions of supergravity. A special class of interpolating solutions, namely solitonic solutions that interpolate between a Lifshitz background and another {\it vacuum} solution, would be an ideal way to holographically model the RG flow from a Lifshitz fixed point in the UV of the dual field theory to a well-behaved IR fixed point. In order to look for such solitonic solutions we performed a first-order rewriting of the one-dimensional action,
\eqref{1d-effaction}, with purely electric gaugings, in terms of a sum of squares. This is a generalization 
of a similar rewriting performed in \cite{Barisch:2011ui} to the case with hypermultiplets. This constitutes an important step 
towards making contact with actual string theory embeddings where hypermultiplets in ${\cal N}=2$ compactifications are omnipresent and ${\cal N}=2$ gauged supergravities (with hypermultiplets) can be obtained in a straightforward manner via flux compactifications. 
Moreover, even though our first-order flow equations \eqref{eomcond} are in complete agreement with the ones obtained in \cite{Halmagyi:2013sla}
via analyzing the supersymmetry conditions,
our rewriting has the advantage to 
allow for a straightforward 
generalization to non-supersymmetric  first-order flows, as briefly discussed at the end of sec.\ \ref{sec:nonsusy}. It would be worthwhile exploring this further. 

The first-order flow equations allow for Lifshitz backgrounds that are characterized by the set of equations given in \eqref{conditions-lif}.
While there is a lower bound on the allowed value of $z$ ($z \geq 1$), the equations do not appear to set an upper bound
on $z$. The explicit examples we constructed have $z=2$, but it would be interesting to see if examples
with $z \neq 1,2$ can be constructed as well. We note that non-supersymmetric flows in four dimensions 
between Lifshitz geometries with different values of $z$
have been considered in \cite{Liu:2012wf,Braviner:2011kz} and a non-supersymmetric solution of a gauged ${\cal N} = 2$ supergravity in four dimensions exhibiting $z \approx 39$ was found in \cite{Donos:2010ax}.

Unfortunately, the charge and flux quantum numbers supporting our Lifshitz solutions do not allow for the usual relativistic vacua $AdS_4$ or $AdS_2 \times \mathbb{R}^2$ in any straightforward way and, thus, we did not obtain any solitonic solutions. Therefore, we henceforth adopted a completely different strategy to find interpolating flows from Lifshitz to a black background which could cloak the singularity at the center of Lifshitz. Generalizing the way in which Lifshitz backgrounds with $z=2$ can be obtained via Scherk-Schwarz reduction from five dimensions (cf.\ \cite{Chemissany:2012du}) led us to write down two classes of four-dimensional non-supersymmetric Nernst  brane solutions, one of them indeed having Lifshitz asymptotics (and the other one having conformal AdS asymptotics).\footnote{Given that the solutions are non-supersymmetric, it would be interesting to perform a stability analysis as performed for example in \cite{Burda:2014jca}.} Like their supersymmetric cousins found in \cite{Barisch:2011ui} they also suffer from a tidal force singularity and diverging scalars. However, these can be regulated by heating them up, as in \cite{Goldstein:2014qha,Dempster:2015xqa}.\footnote{Alternatively, one could analyze how quantum corrections in the ${\cal N}=2$ gauged supergravity action affect the near-horizon geometry, along the lines of \cite{Barisch-Dick:2013xga,Hristov:2016vbm}.} This in turn might help to shed light on the embedding of the Nernst brane in the Hilbert space of the holographically dual field theory. We leave these questions for future work. 

Ultimately, adopting the first strategy of exploring the solution space of the first-order flow equations \eqref{eomcond} should provide us with a classification of extremal 
relativistic and non-relativistic solutions of ${\cal N}=2$ gauged supergravity and, thus, with a systematic understanding of the zero 
temperature ground states of the dual field theories. The second strategy, however, opens up distinct possibilities of constructing a different (non-solitonic) class of interpolating solutions in four dimensional gauged supergravity. It also motivates a thorough exploration of the space of solitonic solutions in the corresponding five-dimensional gauged supergravity which can then be used to generate interpolating four-dimensional solutions via Scherk-Schwarz reduction. 
This avenue of research is currently being pursued in \cite{workinprogress}.


\subsection*{Acknowledgements}

\noindent
We acknowledge helpful discussions with Paolo Benincasa, Harold Erbin, Jelle Hartong, Cynthia Keeler, Olaf Lechtenfeld, Thomas Mohaupt, Niels Obers, Razieh Pourhasan, Larus Thorlacius and 
Marco Zagermann.
The authors are grateful to the Mainz Institute for Theoretical Physics (MITP) for its hospitality and its partial
support during the completion of this work.
The
work of G.L.C. and S. N. was 
supported by FCT/Portugal through UID/MAT/04459/2013, through grant
EXCL/MAT-GEO/0222/2012 and through fellowship SFRH/BPD/101955/2014. 
The work of M.H. was supported by the Excellence Cluster ``The Origin and the Structure of the Universe'' in Munich and by the German Research Foundation (DFG) within the Emmy-Noether-Program (grant number: HA 3448/3-1).
This work was also supported by the COST
action MP1210 {\it ``The String Theory Universe''}. 
G.L.C. and S.N. would like to thank LMU for kind hospitality during various stages of the project. G.L.C. would like to thank the Max-Planck-Institut f\"ur Gravitationsphysik (Albert-Einstein-Institute) for kind hospitality during the completion of this work.

\appendix

\section{${\cal N}=2$ gauged supergravity theories in four dimensions \label{gaugsug}}
\setcounter{equation}{0}

We consider the
bosonic part of the Lagrangian of ${\cal N}=2$ supergravity theories with 
$n_V$ physical vector multiplets and $n_H$ hypermultiplets, in the presence of 
gauging of Abelian isometries of the quaternionic K\"ahler manifold \cite{Andrianopoli:1996vr} (we use the conventions of \cite{Freedman:2012zz}),
\begin{equation}
L = \tfrac12 \, R - N_{IJ} {\cal D}_{\mu} X^I {\cal D}^{\mu} {\bar X}^J 
- \tfrac12 h_{\alpha \beta} D_{\mu} q^{\alpha} D^{\mu} q^{\beta} 
+ \tfrac14 \, {\rm Im} {\cal N}_{IJ}
F^I_{\mu \nu} F^{\mu \nu J} - \tfrac14  {\rm Re} {\cal N}_{IJ} F^I_{\mu \nu} {\tilde F}^{\mu \nu J}
- V(X, \bar X, q) \,.
\label{action4d}
\end{equation}
The complex scalar fields $X^I$ ($I= 0, \dots, n_V$) satisfy the constraint
\begin{equation}
N_{IJ} \, X^I \, {\bar X}^J = -1 \;,
\label{hypersurf}
\end{equation}
where $N_{IJ} = -i (F_{IJ} - {\bar F}_{IJ})$, with $F_{IJ} = \partial^2 F(X) / \partial X^I \partial X^J$.
The holomorphic function $F(X)$ is called the prepotential. The gauge kinetic couplings are given by
\begin{equation}
{\cal N}_{IJ} = {\bar F}_{IJ} + i \, \frac{N_{IK} \, X^K \, N_{JL} \, X^L}{X^M \, N_{MN} \, X^N} \;.
\end{equation}
We note the following useful identity for the inverse of ${\rm Im} \,{\cal N}_{IJ}$,
\begin{equation}
- \tfrac12  ({\rm Im} {\cal N})^{IJ}  = N^{IJ} + {X}^I \bar{{X}}^J + {X}^J \bar{{X}}^I \;.
\label{ImNN}
\end{equation}
The $X^I$ carry a $U(1)$ charge,
\begin{equation} 
{\cal D}_{\mu} X^I = \partial_{\mu} X^I + i {\cal A}_{\mu} X^I \;.
\end{equation}
It will be convenient to introduce $U(1)$ invariant variables ${\tilde X}^I = e^{i \phi} X^I$ by means of a compensating
phase $\phi$, so that
\begin{equation} 
{\cal D}_{\mu} X^I = e^{-i \phi} \left( \partial_{\mu} + i {\tilde {\cal A}}_{\mu} \right) {\tilde X}^I  \;\;\;,\;\;\;
{\tilde {\cal A}}_\mu = {\cal A}_{\mu} - \partial_{\mu} {\phi} \;,
\label{XtilX}
\end{equation}
and
\begin{equation}
N_{IJ} {\cal D}_{\mu} X^I \, {\cal D}^{\mu} {\bar X}^J = 
N_{IJ} {\partial}_{\mu} {\tilde X}^I \, {\partial}^{\mu} {\bar  {\tilde X}}^J + i {\tilde {\cal A}}_{\mu}
\left({\tilde X}^I N_{IJ} \partial^{\mu} {\bar {\tilde X}}^J - {\bar {\tilde X}}^I N_{IJ} \partial^{\mu} {\tilde X}^J \right)
- {\tilde {\cal A}}_{\mu} {\tilde {\cal A}}^{\mu} \;.
\end{equation}
Eliminating ${\tilde {\cal A}}_{\mu}$ using its equation of motion yields ${\tilde {\cal A}}_{\mu} = \Sigma_{\mu}$
with
\begin{equation}
\Sigma_{\mu} = \tfrac12 i 
\left({\tilde X}^I N_{IJ} \partial_{\mu} {\bar {\tilde X}}^J - {\bar {\tilde X}}^I N_{IJ} \partial_{\mu}  {\tilde X}^J \right) = 
i \, {\tilde X}^I N_{IJ} \partial_{\mu} {\bar {\tilde X}}^J = - i {\bar {\tilde X}}^I N_{IJ} \partial_{\mu} {\tilde X}^J
\label{sigmaX}
\end{equation}
by means of \eqref{hypersurf}. Hence
\begin{eqnarray}
N_{IJ} {\cal D}_{\mu} X^I \, {\cal D}^{\mu} {\bar X}^J =
N_{IJ} {\partial}_{\mu} {\tilde X}^I \, {\partial}^{\mu} {\bar {\tilde X}}^J 
+ \Sigma_{\mu} \Sigma^{\mu} \;.
\label{NDXDX}
\end{eqnarray}

The $q^{\alpha}$ ($\alpha = 1, \dots, 4 n_H$) denote local real coordinates of the quaternionic K\"ahler manifold. Gauging its Abelian isometries
results in 
\begin{equation}
D_{\mu} q^{\alpha} = \partial_{\mu} q^{\alpha} + k^{\alpha}_I \, A^I_{\mu} \;,
\end{equation}
and hence
\begin{equation}
h_{\alpha \beta} \, D_{\mu} q^{\alpha} \, D^{\mu} q^{\beta} = h_{\alpha \beta} 
\, \partial_{\mu} q^{\alpha} \, \partial^{\mu} q^{\beta} + 2 h_{\alpha \beta} 
\, \partial_{\mu} q^{\alpha} \, k^{\beta}_I \, A^{I\mu}
+
h_{\alpha \beta} (k^{\alpha}_I \, A^I_{\mu})
(k^{\beta}_J \, A^{J{\mu}}) \;.
\label{kin-quat}
\end{equation}
Here, $k_I = k_I^{\alpha} \, \partial / \partial q^{\alpha}$ denote Killing vectors associated with the isometries
that are being gauged.
The scalar potential $V$ reads (see (21.30) in \cite{Freedman:2012zz})
\begin{equation}
V(X, \bar X, q) = 2 h_{\alpha \beta} (k^{\alpha}_I \,X^I)
(k^{\beta}_J \, {\bar X}^J) + \left( N^{IJ} - 2 X^I {\bar X}^J \right) P^x_I P^x_J \;,
\label{pot}
\end{equation}
where the triplet of Killing prepotentials $P_I^x (q)$ ($x = 1,2,3$) satisfies (see (20.174) in \cite{Freedman:2012zz})
\begin{equation}
\nabla_{\beta} P^x_I = -  k_I^{\alpha} \, \Omega^x_{\alpha \beta} \;,
\label{PkOm}
\end{equation}
with the $SU(2)$ curvature $\Omega^x$ satisfying (see (20.24) in \cite{Freedman:2012zz})
\begin{equation}
\Omega^x_{\alpha \gamma} \, \Omega^{y \; \gamma}{}_{\beta}  = - \delta^{xy} \, h_{\alpha \beta}
+ \epsilon^{x y z} \, \Omega^z_{\alpha \beta} \;.
\label{OmOm}
\end{equation}

\section{First-order rewriting of one-dimensional effective Lagrangian \label{apprew}}
\setcounter{equation}{0}
We consider static electrically charged configurations, specified by the line element\footnote{We reproduce some of
the formulae of the main text for the convenience of the reader.}
\begin{equation}
ds^2 = - {\rm e}^{2 U(r)} \, dt^2 + {\rm e}^{2 A(r)} \left( dr^2 + dx^2 + dy^2 \right) \;,
\end{equation}
and supported by scalar fields $X^I (r), q^{\alpha}(r)$ 
as well as electric fields $E^I (r) = - \partial_r A^I_t (r)$.
For these configurations, we obtain
\begin{eqnarray}
\tfrac12 \sqrt{-g} R &=& - \left[ e^{U + A} (2 A' + U') \right]' + e^{U + A} \left(
(A' + U')^2 - U'^2 \right) \;,
\label{Rconfig}
\end{eqnarray}
where $' = \partial_r$.
Inserting \eqref{Rconfig} into the four-dimensional Lagrangian $(-\sqrt{-g} L)$ given in \eqref{action4d} yields the 
following one-dimensional effective Lagrangian,
\begin{eqnarray}
\label{1d-effaction}
{\cal L}_{1d}&=&  e^{\psi} \left[ U'^2 - \psi'^2 + N_{IJ} {\tilde X}'^I {\bar {\tilde X}}'^J 
+  \Sigma_r^2 
+ \tfrac12 h_{\alpha {\beta}} \, q'^{\alpha} {q}'^{ \beta} 
\right. \\
&& \left. \qquad  - \tfrac12 e^{2(A-U)} \, 
h_{\alpha { \beta}} \,  (k^{\alpha}_I \, A^I_t)
(k^{{\beta}}_J \, A^J_t) 
+ \tfrac12 \, e^{-2U} \, {\rm Im} {\cal N}_{IJ} \, E^I E^J + e^{2A} V({\tilde X}, \bar{\tilde X}, q)  \right] + {\rm T.D.} \;, \nonumber
\end{eqnarray}
where we introduced $\psi = A + U$
for convenience, and where
${\rm T.D.}$ denotes a total derivative, 
\begin{equation}
{\rm T.D.}  =  \left[ e^{\psi} (2 \psi' - U') \right]' \;.
\end{equation}
In obtaining \eqref{1d-effaction}, we used \eqref{NDXDX} and \eqref{kin-quat}, restricted to electric gaugings. Moreover, $\Sigma_r$ is given in \eqref{sigmaX}.

It will be convenient to work with 
rescaled scalar fields $Y^I = e^A \, {\tilde X}^I$. Using \eqref{hypersurf} we obtain
\begin{eqnarray}
e^{2A} &=& - N_{IJ} \, Y^I \, {\bar Y}^J \;, \nonumber \\
N_{IJ} {\tilde X}'^I {\bar {\tilde X}}'^J &=& e^{-2A} \, N_{IJ} Y'^I {\bar Y}'^J + A'^2 \;.
\label{AYY}
\end{eqnarray}
We also introduce a new set of real fields $C_I$ and rewrite
\begin{eqnarray} 
 \tfrac12 \, e^{-2U} \, {\rm Im} {\cal N}_{IJ} \, E^I E^J &=& 
  \tfrac12 \, e^{-2U} \, {\rm Im} {\cal N}_{IJ} \, \left( E^I + e^{U + A} ({\rm Im} {\cal N})^{IK} \, C_K \right)
   \left( E^J + e^{U + A} ({\rm Im} {\cal N})^{JL} \, C_L \right) \nonumber\\
   && -  \tfrac12 \, e^{2A} \, ({\rm Im} {\cal N})^{IJ} \, C_I C_J - e^{A-U} C_I E^I \;,
   \label{E-BPS}
\end{eqnarray}
where $ ({\rm Im} {\cal N})^{IJ}$ denotes the inverse of $ ({\rm Im} {\cal N})_{IJ}$. Below we will use the identity
\eqref{ImNN}
to rewrite terms involving $ ({\rm Im} {\cal N})^{IJ}$, such as $({\rm Im} {\cal N})^{IJ} \, C_I C_J $.

Next, we introduce the combination
\begin{equation}
q_I = e^{i \gamma(r)} \left(C_I - i P^3_I \right) \;,
\label{def-qI1}
\end{equation}
where $\gamma$ is an $r$-dependent phase and $P^3_I$ denotes one of the three real Killing prepotentials
appearing in the potential $V$ given in \eqref{pot}. 
Using \eqref{def-qI1}, we rewrite
\begin{eqnarray}
 e^{-2A} \, N_{IJ} Y'^I {\bar Y}'^J &=& e^{-2A} \, N_{IJ} \left( Y'^I  - e^{2A} N^{IK} {\bar q}_K \right) 
 \left( {\bar Y}'^J - e^{2A} N^{JL} q_L \right) - e^{2A} q_I N^{IJ} {\bar q}_J \nonumber\\
 &&+ 2 \, {\rm Re} \left(Y'^I \, q_I \right) \;.
 \label{Y-BPS}
\end{eqnarray}
Next, performing partial integrations on the last terms of \eqref{E-BPS} and \eqref{Y-BPS}
we obtain
\begin{eqnarray}
e^{\psi} \left[- e^{A-U} C_I E^I + 2  \, {\rm Re} \left(Y'^I \, q_I \right) \right]
&=& \left[ e^{\psi + A - U} \, A^I_t C_I + 2 e^{\psi }  \, {\rm Re} \left(Y^I \, q_I \right)\right]'
\\
&& - e^{\psi} \left[ 2 \psi' \,   {\rm Re} \left( Y^I q_I\right) 
+ 2 {\rm Re} \left( Y^I q_I' \right) \right. \nonumber\\
&&\qquad  \left. + (\psi' + A' - U') e^{A-U} A_t^I C_I + e^{A-U} C_I' A_t^I \right] \,. \nonumber
\end{eqnarray}
Using $\psi'  = 2 \psi' - U' - A'$ and $\psi' + A' - U' = 2(\psi' - U')$ we obtain
\begin{eqnarray}
e^{\psi} \left[- e^{A-U} C_I E^I + 2  \, {\rm Re} \left(Y'^I \, q_I \right) \right]
&=& {\rm T. D.}
 - e^{\psi} \left[ 2 (2\psi' - U' - A')  {\rm Re} \left( Y^I q_I \right) \right. \nonumber\\
&& \qquad - 2  \gamma'  \left( {\rm Im} \left( e^{i \gamma}
Y^I C_I   \right) - {\rm Re} \left(e^{i \gamma}
Y^I P^3_I  \right) \right)\nonumber\\
&& \left. \qquad 
+ 2 {\rm Re} \left( e^{i \gamma} \, Y^I C_I' \right) 
+ 2 {\rm Im} \left( e^{i \gamma} \, Y^I \right) (P^3_I)' \right. \nonumber\\
&&\qquad  \left. + 2 (\psi' - U') e^{A-U} A_t^I C_I + e^{A-U} C_I' A_t^I \right] \,.
\label{C'W'}
\end{eqnarray}
Collecting all the terms involving derivatives of $U, \psi$ and $A$, we obtain
\begin{eqnarray}
&& U'^2 - \psi'^2 + A'^2 -  2 (2\psi' - U' - A')  {\rm Re} \left( Y^I q_I \right) 
-2 (\psi' - U') e^{A-U} A_t^I C_I = \nonumber\\
&& \left(U' + {\rm Re} \left( Y^I q_I \right) + e^{A-U} A_t^I C_I \right)^2 
 -  \left(\psi' + 2 {\rm Re} \left( Y^I q_I \right) + e^{A-U} A_t^I C_I \right)^2 
 + \left(A' + {\rm Re} \left( Y^I q_I \right)\right)^2 \nonumber\\
&& + 2  \left( {\rm Re} \left( Y^I q_I \right)\right)^2 + 2 e^{A-U} A_t^I C_I \,
\, {\rm Re} \left( Y^J q_J \right) \;.
\end{eqnarray}

Next, combining the term proportional to $(P^3_I)'$ in \eqref{C'W'}, 
 with the kinetic term for the hypermultiplet
scalar fields in \eqref{1d-effaction} yields 
\begin{eqnarray}
&& \tfrac12 h_{\alpha \beta}    \,  q'^{\alpha} { q}'^{ \beta}  - 2  \, {\rm Im} \left( e^{i \gamma} \, Y^I  \right) 
q'^{\gamma}\,  \nabla_\gamma P^3_I 
 =
\nonumber\\
 && \tfrac12 h_{{\alpha} {\beta}} \left( q'^{\alpha} - 2  \, h^{\alpha {\gamma}} \,
 {\rm Im} \left( e^{i \gamma} \, Y^I  \right)
 \nabla_\gamma P^3_I 
  \right)
  \left( {q}'^{{\beta}} - 2  \, h^{\beta \delta  }\,
 {\rm Im} \left( e^{i \gamma} \, Y^J  \right)
 \nabla_\delta P^3_J 
  \right)
  \nonumber\\
   && - 2 \, 
  {\rm Im} \left( e^{i \gamma} \, Y^I \right) {\rm Im} \left( e^{i \gamma} \, Y^J \right)
 \nabla_\alpha P^3_I \, h^{\alpha {\beta}}   \, 
\nabla_\beta P^3_J  \;.
   \end{eqnarray}

Thus, the one-dimensional effective Lagrangian \eqref{1d-effaction}  can be written as
\begin{equation}
{\cal L}_{1d} = {\rm T.D.} + {\cal L}_{\rm squares} + \Delta \;,
\label{1d-effaction_rewrite2}
\end{equation}
where
$\rm T.D.$ denotes a total derivative term, 
${\cal L}_{\rm squares}$ denotes a  sum of squares, 
\begin{eqnarray}
{\cal L}_{\rm squares} &=& e^{\psi} \left[
\left(U' +  {\rm Re} \left( Y^I q_I \right) + e^{A-U} A_t^I C_I \right)^2 
 -  \left(\psi' + 2  {\rm Re} \left( Y^I q_I \right) + e^{A-U} A_t^I C_I \right)^2 \right.
 \nonumber\\
&&\quad  \left. + \left(A' +  {\rm Re} \left( Y^I q_I \right)\right)^2
  + e^{-2A} \, N_{IJ} \left( Y'^I  - e^{2A} N^{IK} {\bar q}_K \right) 
 \left( {\bar Y}'^J - e^{2A} N^{JL} q_L \right) 
+ \Sigma_r^2 \right. \nonumber\\
&&\quad  
 + \tfrac12 h_{{\alpha} {\beta}} \left( q'^{\alpha} - 2 \, h^{\alpha {\gamma}} \,
 {\rm Im} \left( e^{i \gamma} \, Y^I  \right)
\nabla_\gamma P^3_I 
  \right)
  \left( q'^{{\beta}} - 2 \, h^{\beta \delta }\,
 {\rm Im} \left( e^{i \gamma} \, Y^J  \right)
\nabla_\delta P^3_J
  \right)
  \nonumber\\
  && \quad \left. +  \tfrac12 \, e^{-2U} \, {\rm Im} {\cal N}_{IJ} \, \left( E^I + e^{U + A} ({\rm Im} {\cal N})^{IK} \, C_K \right)
   \left( E^J + e^{U + A} ({\rm Im} {\cal N})^{JL} \, C_L \right) 
   \right] \;, \nonumber\\
\end{eqnarray}
while $\Delta$ is given by 
\begin{eqnarray}
\Delta &=& e^{\psi} \left[ 2 \Lambda^- \left( \gamma' +  \Lambda^+ \right) - \tfrac12 
e^{2(A-U)}  \, h_{\alpha {\beta}} \,
(k^{\alpha}_I {\tilde A}_t^I)  \, (k^{ \beta}_J {\tilde A}_t^J) - e^{A - U} \, C_I' \, {\tilde A}^I_t 
 \right. \nonumber \\ 
&& \qquad  + 2 e^{A-U} \left(  h_{\alpha {\beta}} \, ( k_I^{\alpha}  {\tilde A}_t^I ) \,  k_J^{ \beta}   \, {\rm Re} (Y^J e^{i \gamma}) 
+ C_I {\tilde A}_t^I \, {\rm Re} (q_J Y^J)
\right) \nonumber\\
&& \left. \qquad  + 2 {\rm Im} \left( e^{i \gamma} \, Y^I \right) {\rm Im} \left( e^{i \gamma} \, Y^J \right)
\left( k_I^{\alpha} \, h_{\alpha {\beta}} \,  k_J ^{ \beta} -
 \nabla_\alpha P^3_I \, h^{\alpha {\beta}}   \, 
\nabla_\beta P^3_J \right)
   \right. 
   \nonumber\\
   && \qquad  
   \left. +  \left(e^{2A} \,  N^{IJ} - 2 Y^I {\bar Y}^J \right) \left(P^1_I P^1_J + P^2_I P^2_J \right) 
  \right]
 \;,
 \label{Delta}
\end{eqnarray}
where we introduced the combinations
\begin{eqnarray}
{\tilde A}_t^I &=& A_t^I + 2 e^{U-A} \, {\rm Re} \left(Y^I e^{i \gamma} \right) \;, \nonumber\\
\Lambda^{\pm} &=& {\rm Im} \left( C_I Y^I e^{i \gamma} \right) \pm {\rm Re} \left(
P^3_I Y^I e^{i \gamma} \right) \;.
\end{eqnarray}
On the quaternionic K\"ahler manifold, the relations \eqref{PkOm} and \eqref{OmOm} imply the identity
\begin{equation}
 \nabla_\alpha P^3_I \, h^{\alpha {\beta}}   \, 
   \nabla_\beta P^3_J = \, k_I^{\alpha} \, h_{\alpha \beta} \, k_J^{\beta} \;.
   \label{relPk}
   \end{equation}
This eliminates the third line of $\Delta$ in \eqref{Delta}, leading to
\begin{eqnarray}
\Delta &=& e^{\psi} \left[ 2 \Lambda^- \left( \gamma' +  \Lambda^+ \right) - \tfrac12 
e^{2(A-U)}  \, h_{\alpha {\beta}} \,
(k^{\alpha}_I {\tilde A}_t^I)  \, (k^{ \beta}_J {\tilde A}_t^J) - e^{A - U} \, C_I' \, {\tilde A}^I_t 
 \right. \nonumber \\ 
&& \qquad  + 2 e^{A-U} \left(  h_{\alpha {\beta}} \, ( k_I^{\alpha}  {\tilde A}_t^I ) \,  k_J^{ \beta}   \, {\rm Re} (Y^J e^{i \gamma}) 
+ C_I {\tilde A}_t^I \, {\rm Re} (q_J Y^J)
\right) \nonumber\\
   && \qquad  
   \left. +  \left(e^{2A} \,  N^{IJ} - 2 Y^I {\bar Y}^J \right) \left(P^1_I P^1_J + P^2_I P^2_J \right) 
  \right]
 \;.
\end{eqnarray}

\section{Some details concerning adding magnetic charge densities \label{sec:details_magnetic_charges} }  
\setcounter{equation}{0}

We begin by recalling how to incorporate magnetic charge densities $p^I$ in the absence of mass terms for the
Abelian gauge fields.  We will denote the electric charges by $Q_I$ for the time being.

Let us consider the Maxwell part of the Lagrangian \eqref{action4d},
\begin{equation}
L_{\rm Maxwell}
=  \tfrac14 \, {\rm Im} {\cal N}_{IJ}
F^I_{\mu \nu} F^{\mu \nu J} - \tfrac14 {\rm Re} {\cal N}_{IJ} F^I_{\mu \nu} {\tilde F}^{\mu \nu J} \;.
\end{equation}
Consider non-vanishing electric fields $E^I = F^I_{tr}$ and constant magnetic fields
$F^I_{xy} = p^I$.  Inserting these into $L_{\rm Maxwell}$ yields
\begin{equation}
- \sqrt{-g} \, L_{\rm Maxwell}
= \tfrac12 e^{A-U} \left[ E^I \, \mu_{IJ} \, E^J - p^I \, \mu_{IJ} \, p^J \, e^{2(U-A)} + 2 E^I \, \nu_{IJ} \, p^J \, e^{U-A}
\right]\ ,
\end{equation}
where $\mu_{IJ} =  {\rm Im} {\cal N}_{IJ}$ and $\nu_{IJ} =  {\rm Re} {\cal N}_{IJ}$. Adding and subtracting a total
derivative term $Q_I E^I$ yields
\begin{eqnarray} \label{Maxwell_magnetic}
- \sqrt{-g} \, L_{\rm Maxwell}
&=& \tfrac12 e^{A-U} \left[ \left(E + \mu^{-1} (Q + \nu \, p)  \, e^{U-A}   \right)^T \, \mu \, 
\left(E + \mu^{-1} (Q + \nu \, p)  \, e^{U-A}    \right) \right. \nonumber\\
&& \left. -  \left(Q + \nu \, p \right)^T \mu^{-1} \left(Q + \nu \, p \right) \, e^{2(U-A)} - p^T  \mu\, p \, e^{2(U-A)} 
\right] - Q_I E^I \nonumber\\
&=& \tfrac12 e^{A-U} \, \left(E + \mu^{-1} (Q + \nu \, p ) \, e^{U-A}       \right)^T \, \mu \, 
\left(E + \mu^{-1} (Q + \nu \, p)  \, e^{U-A}      \right)  \nonumber\\
&& + e^{U-A} \, V_{\rm BH}  - Q_I E^I \;,
\end{eqnarray}
with the potential $V_{\rm BH}$ given by
\begin{eqnarray}
V_{\rm BH} = -  \tfrac12 
\begin{pmatrix}
p & Q \\
\end{pmatrix}
\begin{pmatrix}
\mu + \nu \mu^{-1} \nu & \nu \mu^{-1} \\
\mu^{-1} \nu  & \mu^{-1}\\
\end{pmatrix}
\begin{pmatrix}
p \\
 Q \\
\end{pmatrix} \;.
\end{eqnarray}
This can also be expressed as 
\begin{equation}
V_{\rm BH} = g^{i \bar{\jmath}} {\cal D}_i Z \bar{\cal D}_{\bar \jmath} {\bar Z} + |Z|^2 = 
\left( N^{IJ} + 2 {X}^I \bar{{X}}^J \right) {\hat Q}_I  \, \bar{\hat Q}_J \;,
\end{equation}
where
\begin{equation}
Z = - \hat{Q}_I \, X^I \;\;\;,\;\;\; \hat{Q}_I = Q_I - F_{IJ} \, p^J \;.
\end{equation}

Now let us return to the models considered here, where we used $C_I$ rather than $Q_I$. Both are
related by  $Q_I = e^{2A} \, C_I$, and hence 
we may readily incorporate magnetic charge densities by replacing
$C_I$ in \eqref{def-qI} with
\begin{equation}
\hat{C}_I = C_I - F_{IJ}  \, p^J \, e^{-2A} \;, 
\end{equation}
as discussed in the main text, cf.\ \eqref{Chat}. From \eqref{Maxwell_magnetic} we also infer the form of the electric field given in \eqref{EI}.

\section{Parametrizations of the coset space $SU(2,1)/SU(2) \times U(1)$ \label{quater-metric} }  
\setcounter{equation}{0}

We use the following two standard parametrizations of the coset space $SU(2,1)/SU(2) \times U(1)$ 
in terms of complex coordinates $q^i \, (i=1,2)$ \cite{Halmagyi:2011xh}.
In the first parametrization, the coset metric is given by 
\begin{equation}
ds^2 = h_{i \bar \jmath} \, dq^i \, d {\bar q}^{\bar \jmath} = 
\frac{dq^{1} d {\bar q}^{\bar 1} + dq^{2} d {\bar q}^{\bar 2}
}{1 - |q^1|^2 - |q^2|^2 }
+
\frac{\left(q^1 d {\bar q}^{\bar 1} + q^2 d {\bar q}^{\bar 2} \right) \left({\bar q}^{\bar 1} dq^1 + {\bar q}^{\bar 2}
d q^2\right)}{
\left(1 - |q^1|^2 - |q^2|^2 \right)^2} \;.
\label{quat-metric}
\end{equation}
The second parametrization uses the K\"ahler potential
\begin{equation}
K  = - \ln \left( \frac12 \left(S + \bar S - 2 \xi {\bar \xi} \right) \right) \;.
\end{equation}
The resulting metric reads (with $q^i = S, \xi$) 
\begin{eqnarray}
ds^2 &=& h_{i \bar \jmath} \, d q^i d \bar{q}^{\bar \jmath}
= \frac{1}{\left(S + \bar S - 2 \xi {\bar \xi} \right) ^2} \left[ dS d \bar S - 2 \xi \, dS d \bar \xi - 2 {\bar \xi} \, d \xi d \bar S + 2 (S + \bar S) \, d \xi d \bar \xi \right] \;.
\nonumber\\
\label{cosetmet}
\end{eqnarray}
Introducing real coordinates $S = D + i \sigma$ and $\xi = \chi \, e^{i \theta}$, the
metric becomes
\begin{eqnarray}
ds^2 = \frac {1}{4 \rho^2} d \rho^2 + \frac{1}{4 \rho^2} \left( d \sigma - i (\xi d {\bar \xi} - {\bar \xi} d \xi ) \right)^2 + \frac{1}{\rho} d \xi  d {\bar \xi} \;,
\label{met-pet}
\end{eqnarray}
where $\rho = D - \chi^2$. 


\section{Proof that the solutions of sec.\ \ref{sec:interpolating} do not solve the first-order flow equations \label{sec:no_solution_flow_eqs}}  
\setcounter{equation}{0}

First of all, it is easy to see that the solution  \eqref{deform-line} -- \eqref{f} does not solve the first-order flow equations \eqref{eomcond}. It suffices to consider the equations for $Y^I$. The first equation in \eqref{AYY} fixes the $Y^I$ up to an overall phase, which in general could be $r$-dependent, i.e.\  
\bea
&& \hspace{-0.5cm} Y^0 = -\frac{1}{2 \sqrt{2}} \frac{\sqrt{b_1 b_2 b_3}}{\sqrt{r^2+\alpha r^4+\beta} \,  C_0^{3/2}} e^{i \delta(r)}\ ,  \label{Y0} \\
&& \hspace{-0.5cm} Y^1 = -\frac{i}{2 \sqrt{2}} \frac{\sqrt{b_2 b_3}}{\sqrt{b_1 C_0} r} e^{i \delta(r)}\ , \quad Y^2 = -\frac{i}{2 \sqrt{2}} \frac{\sqrt{b_1 b_3}}{\sqrt{b_2 C_0} r} e^{i \delta(r)}\ , \quad Y^3 = -\frac{i}{2 \sqrt{2}} \frac{\sqrt{b_1 b_2}}{\sqrt{b_3 C_0} r} e^{i \delta(r)}. \label{YA}
\eea
The overall $r$-dependent phase in $Y^I$ drops out in the physical scalar fields $z^A= Y^A/Y^0$. 

Plugging \eqref{Y0} into the $Y^0$ equation in \eqref{eomcond} and \eqref{YA} into the $Y^A$ equations, one obtains
\bea
&& -\frac{i (\alpha r^4+r^2+\beta) \delta'(r) - 2 \alpha r^3-r}{(\alpha r^4+r^2+\beta)^{1/2}} e^{i \delta(r)} = e^{-i \gamma(r)}\ , \label{Y0eq} \\
&&  (-i \delta'(r) r+1) e^{i \delta(r)} = e^{-i \gamma(r)}\ . \label{YAeq}
\eea
This implies
\be \label{deltaprime}
\delta'(r) = i \frac{-2 \alpha r^3-r+\sqrt{\alpha r^4+r^2+\beta}}{\alpha r^4+ r^2+\beta-\sqrt{\alpha r^4+r^2+\beta} r}\ .
\ee
This does not have any real solution for non-vanishing $\alpha$ and/or $\beta$.\footnote{Obviously, for $\alpha = \beta = 0$, \eqref{Y0eq} and \eqref{YAeq} are solved by $\delta = \gamma = 0$.}

This state of affairs does not change when introducing a matrix $M$ as in the discussion at the end of sec.\ \ref{electric_rewrite}. To see this, let us look at the $Y^I$ equations of \eqref{eomcond} again, i.e.
\be \label{Yeq}
Y'^I = e^{2A} N^{IK} \bar{q}_K
\ee
with $N^{IK}$ given in \eqref{NIJmatrix} and 
\be
\bar{q}_K = e^{-i \gamma(r)} [((M^{-1})^T C)_K + i P_K^3] = e^{-i \gamma(r)} [((M^{-1})^T)_K\ \!\! ^0 C_0 + i P_K^3] \equiv e^{-i \gamma(r)} [t_K + i P_K^3]  \ ,
\ee
where $P_0^3=0$, $P_A^3 = -b_A$ and $t_K$ are real. 
The equations for $Y^0$ and $Y^A$ become more complicated but one can go through similar steps as before for the $Y^0$ and $Y^1$ equations in order to obtain (for $b_1 = b_2 = 1$ and $b_3=2$ for concreteness)
\be \label{deltaprime}
\delta'(r) = \frac{-2i \sqrt{\alpha r^4+r^2+\beta} t_0 + 4(\alpha r^2+\tfrac12) (i+t_1-t_2-\tfrac{t_3}{2}) r}{2 \sqrt{\alpha r^4+r^2+\beta} t_0 r + 2 (i t_1- i t_2 - i \tfrac{t_3}{2} - 1) (\alpha r^4+r^2+\beta)}\ .
\ee
If one instead takes the $Y^0$ and $Y^2$ or $Y^0$ and $Y^3$ equations one obtains similar equations just with the combination $t_1-t_2-\tfrac{t_3}{2}$ replaced by $-t_1+t_2-\tfrac{t_3}{2}$ and $-t_1-t_2+\tfrac{t_3}{2}$, respectively. Thus, in order to get a unique equation for $\delta'$, one has to have the relations $t_1=t_2$ and $t_3=2 t_2$ (the asymmetry between $t_1, t_2$ and $t_3$ arises from our asymmetric choice of the $b_A$). Substituting this into \eqref{deltaprime} gives
\be 
\delta'(r) = \frac{-2i \sqrt{\alpha r^4+r^2+\beta} t_0 + 4(\alpha r^2+\tfrac12) (i-t_1) r}{2 \sqrt{\alpha r^4+r^2+\beta} t_0 r - 2 (i t_1 + 1) (\alpha r^4+r^2+\beta)}\ .
\ee
The right hand side has a non-trivial imaginary part given by
\be \label{imdelta}
\frac{-3\Big(\alpha r^4+\tfrac23 r^2+\tfrac13 \beta \Big) t_0 \sqrt{\alpha r^4+r^2+\beta}+2 r \Big(\alpha (t_1^2+1) r^2+\tfrac12 t_0^2+\tfrac12 t_1^2+\tfrac12\Big) (\alpha r^4+r^2+\beta)}{\Big(-\alpha r^4 t_1^2-\alpha r^4-t_0^2 r^2-r^2 t_1^2+2 \sqrt{\alpha r^4+r^2+\beta} t_0 r-t_1^2 \beta-r^2-\beta\Big) (\alpha r^4+r^2+\beta)}\ .
\ee
This is non-trivial for arbitrary choices of $t_0$ and $t_1$ if $\alpha$ and/or $\beta$ are non-vanishing. For instance, choosing $\alpha = 1$ and $\beta=0$, \eqref{imdelta} becomes
\be
\frac{(-3 r^2 t_0 - 2 t_0) \sqrt{r^4+r^2} + 2 r (r^2+1) \Big((t_1^2+1) r^2 + \tfrac12 t_0^2 + \tfrac12 t_1^2 + \tfrac12 \Big)}{r (r^2+1) (-r^3 t_1^2 - r^3 - r t_0^2 - r t_1^2 + 2 t_0 \sqrt{r^4+r^2}-r)}
\ee
There is no choice for $t_0$ and $t_1$ for which this vanishes and, thus, for which $\delta$ is purely real. 


\bibliographystyle{JHEP}
\bibliography{references}

\begin{thebibliography}{10}

\bibitem{Maldacena:1997re}
J.~M. Maldacena, {\it {The Large N limit of superconformal field theories and
  supergravity}},  {\em Int. J. Theor. Phys.} {\bf 38} (1999) 1113--1133,
  [\href{http://xxx.lanl.gov/abs/hep-th/9711200}{{\tt hep-th/9711200}}]. [Adv.
  Theor. Math. Phys. 2, 231 (1998)].

\bibitem{Gubser:1998bc}
S.~S. Gubser, I.~R. Klebanov, and A.~M. Polyakov, {\it {Gauge theory
  correlators from noncritical string theory}},  {\em Phys. Lett.} {\bf B428}
  (1998) 105--114, [\href{http://xxx.lanl.gov/abs/hep-th/9802109}{{\tt
  hep-th/9802109}}].

\bibitem{Witten:1998qj}
E.~Witten, {\it {Anti-de Sitter space and holography}},  {\em Adv. Theor. Math.
  Phys.} {\bf 2} (1998) 253--291,
  [\href{http://xxx.lanl.gov/abs/hep-th/9802150}{{\tt hep-th/9802150}}].

\bibitem{Hartnoll:2009sz}
S.~A. Hartnoll, {\it {Lectures on holographic methods for condensed matter
  physics}},  {\em Class. Quant. Grav.} {\bf 26} (2009) 224002,
  [\href{http://xxx.lanl.gov/abs/0903.3246}{{\tt 0903.3246}}].

\bibitem{McGreevy:2009xe}
J.~McGreevy, {\it {Holographic duality with a view toward many-body physics}},
  {\em Adv. High Energy Phys.} {\bf 2010} (2010) 723105,
  [\href{http://xxx.lanl.gov/abs/0909.0518}{{\tt 0909.0518}}].

\bibitem{Hartnoll:2011fn}
S.~A. Hartnoll, {\it {Horizons, holography and condensed matter}},
  \href{http://xxx.lanl.gov/abs/1106.4324}{{\tt 1106.4324}}.

\bibitem{Barisch:2011ui}
S.~Barisch, G.~L. Cardoso, M.~Haack, S.~Nampuri, and N.~A. Obers, {\it {Nernst
  branes in gauged supergravity}},  {\em JHEP} {\bf 11} (2011) 090,
  [\href{http://xxx.lanl.gov/abs/1108.0296}{{\tt 1108.0296}}].

\bibitem{Dempster:2015xqa}
P.~Dempster, D.~Errington, and T.~Mohaupt, {\it {Nernst branes from special
  geometry}},  {\em JHEP} {\bf 05} (2015) 079,
  [\href{http://xxx.lanl.gov/abs/1501.07863}{{\tt 1501.07863}}].

\bibitem{Son:2007}
 Y.~Nishida and D.~T.~Son,
 {\it {Nonrelativistic conformal field theories}},
 {\em Phys. Rev.} {\bf D76} (2007) 086004,
 [\href{http://xxx.lanl.gov/abs/0706.3746}{{\tt 0706.3746}}].

\bibitem{Kachru:2008yh}
S.~Kachru, X.~Liu, and M.~Mulligan, {\it {Gravity duals of Lifshitz-like fixed
  points}},  {\em Phys. Rev.} {\bf D78} (2008) 106005,
  [\href{http://xxx.lanl.gov/abs/0808.1725}{{\tt 0808.1725}}].

\bibitem{Charmousis:2010zz}
C.~Charmousis, B.~Gouteraux, B.~S. Kim, E.~Kiritsis, and R.~Meyer, {\it
  {Effective Holographic Theories for low-temperature condensed matter
  systems}},  {\em JHEP} {\bf 11} (2010) 151,
  [\href{http://xxx.lanl.gov/abs/1005.4690}{{\tt 1005.4690}}].

\bibitem{Copsey:2010ya}
K.~Copsey and R.~Mann, {\it {Pathologies in Asymptotically Lifshitz
  Spacetimes}},  {\em JHEP} {\bf 03} (2011) 039,
  [\href{http://xxx.lanl.gov/abs/1011.3502}{{\tt 1011.3502}}].

\bibitem{Horowitz:2011gh}
G.~T. Horowitz and B.~Way, {\it {Lifshitz Singularities}},  {\em Phys. Rev.}
  {\bf D85} (2012) 046008, [\href{http://xxx.lanl.gov/abs/1111.1243}{{\tt
  1111.1243}}].

\bibitem{Hartong:2014pma}
J.~Hartong, E.~Kiritsis, and N.~A. Obers, {\it {Schroedinger Invariance from
  Lifshitz Isometries in Holography and Field Theory}},  {\em Phys. Rev.} {\bf
  D92} (2015) 066003, [\href{http://xxx.lanl.gov/abs/1409.1522}{{\tt
  1409.1522}}].
  
\bibitem{Blau:2009gd}
M.~Blau, J.~Hartong, and B.~Rollier, {\it {Geometry of Schrodinger Space-Times,
  Global Coordinates, and Harmonic Trapping}},  {\em JHEP} {\bf 07} (2009) 027,
  [\href{http://xxx.lanl.gov/abs/0904.3304}{{\tt 0904.3304}}].

\bibitem{Balasubramanian:2010uk}
K.~Balasubramanian and K.~Narayan, {\it {Lifshitz spacetimes from AdS null and
  cosmological solutions}},  {\em JHEP} {\bf 08} (2010) 014,
  [\href{http://xxx.lanl.gov/abs/1005.3291}{{\tt 1005.3291}}].

\bibitem{Chemissany:2012du}
W.~Chemissany, D.~Geissbuhler, J.~Hartong, and B.~Rollier, {\it {Holographic
  Renormalization for z=2 Lifshitz Space-Times from AdS}},  {\em Class. Quant.
  Grav.} {\bf 29} (2012) 235017, [\href{http://xxx.lanl.gov/abs/1205.5777}{{\tt
  1205.5777}}].

\bibitem{Ferrara:1996dd}
S.~Ferrara and R.~Kallosh, {\it {Supersymmetry and attractors}},  {\em Phys.
  Rev.} {\bf D54} (1996) 1514--1524,
  [\href{http://xxx.lanl.gov/abs/hep-th/9602136}{{\tt hep-th/9602136}}].

\bibitem{Ferrara:1996um}
S.~Ferrara and R.~Kallosh, {\it {Universality of supersymmetric attractors}},
  {\em Phys. Rev.} {\bf D54} (1996) 1525--1534,
  [\href{http://xxx.lanl.gov/abs/hep-th/9603090}{{\tt hep-th/9603090}}].

\bibitem{Ferrara:1997tw}
S.~Ferrara, G.~W. Gibbons, and R.~Kallosh, {\it {Black holes and critical
  points in moduli space}},  {\em Nucl. Phys.} {\bf B500} (1997) 75--93,
  [\href{http://xxx.lanl.gov/abs/hep-th/9702103}{{\tt hep-th/9702103}}].

\bibitem{Goldstein:2005hq}
K.~Goldstein, N.~Iizuka, R.~P. Jena, and S.~P. Trivedi, {\it
  {Non-supersymmetric attractors}},  {\em Phys. Rev.} {\bf D72} (2005) 124021,
  [\href{http://xxx.lanl.gov/abs/hep-th/0507096}{{\tt hep-th/0507096}}].

\bibitem{Cacciatori:2009iz}
S.~L. Cacciatori and D.~Klemm, {\it {Supersymmetric AdS(4) black holes and
  attractors}},  {\em JHEP} {\bf 01} (2010) 085,
  [\href{http://xxx.lanl.gov/abs/0911.4926}{{\tt 0911.4926}}].

\bibitem{Dall'Agata:2010gj}
G.~Dall'Agata and A.~Gnecchi, {\it {Flow equations and attractors for black
  holes in N = 2 U(1) gauged supergravity}},  {\em JHEP} {\bf 03} (2011) 037,
  [\href{http://xxx.lanl.gov/abs/1012.3756}{{\tt 1012.3756}}].

\bibitem{Hristov:2010ri}
K.~Hristov and S.~Vandoren, {\it {Static supersymmetric black holes in $AdS_4$
  with spherical symmetry}},  {\em JHEP} {\bf 04} (2011) 047,
  [\href{http://xxx.lanl.gov/abs/1012.4314}{{\tt 1012.4314}}].

\bibitem{Katmadas:2014faa}
S.~Katmadas, {\it {Static BPS black holes in U(1) gauged supergravity}},  {\em
  JHEP} {\bf 09} (2014) 027, [\href{http://xxx.lanl.gov/abs/1405.4901}{{\tt
  1405.4901}}].

\bibitem{Son:2008ye}
D.~T. Son, {\it {Toward an AdS/cold atoms correspondence: A Geometric
  realization of the Schrodinger symmetry}},  {\em Phys. Rev.} {\bf D78} (2008)
  046003, [\href{http://xxx.lanl.gov/abs/0804.3972}{{\tt 0804.3972}}].

\bibitem{Balasubramanian:2008dm}
K.~Balasubramanian and J.~McGreevy, {\it {Gravity duals for non-relativistic
  CFTs}},  {\em Phys. Rev. Lett.} {\bf 101} (2008) 061601,
  [\href{http://xxx.lanl.gov/abs/0804.4053}{{\tt 0804.4053}}].

\bibitem{Cassani:2011sv}
D.~Cassani and A.~F. Faedo, {\it {Constructing Lifshitz solutions from AdS}},
  {\em JHEP} {\bf 05} (2011) 013,
  [\href{http://xxx.lanl.gov/abs/1102.5344}{{\tt 1102.5344}}].

\bibitem{Halmagyi:2011xh}
N.~Halmagyi, M.~Petrini, and A.~Zaffaroni, {\it {Non-Relativistic Solutions of
  N=2 Gauged Supergravity}},  {\em JHEP} {\bf 08} (2011) 041,
  [\href{http://xxx.lanl.gov/abs/1102.5740}{{\tt 1102.5740}}].

\bibitem{Chimento:2015rra}
S.~Chimento, D.~Klemm, and N.~Petri, {\it {Supersymmetric black holes and
  attractors in gauged supergravity with hypermultiplets}},  {\em JHEP} {\bf
  06} (2015) 150, [\href{http://xxx.lanl.gov/abs/1503.09055}{{\tt
  1503.09055}}].

\bibitem{Halmagyi:2013sla}
N.~Halmagyi, M.~Petrini, and A.~Zaffaroni, {\it {BPS black holes in $AdS_{4}$
  from M-theory}},  {\em JHEP} {\bf 08} (2013) 124,
  [\href{http://xxx.lanl.gov/abs/1305.0730}{{\tt 1305.0730}}].

\bibitem{Chemissany:2011mb}
W.~Chemissany and J.~Hartong, {\it {From D3-Branes to Lifshitz Space-Times}},
  {\em Class. Quant. Grav.} {\bf 28} (2011) 195011,
  [\href{http://xxx.lanl.gov/abs/1105.0612}{{\tt 1105.0612}}].

\bibitem{Goldstein:2014qha}
K.~Goldstein, S.~Nampuri, and A.~Veliz-Osorio, {\it {Heating up branes in
  gauged supergravity}},  {\em JHEP} {\bf 08} (2014) 151,
  [\href{http://xxx.lanl.gov/abs/1406.2937}{{\tt 1406.2937}}].

\bibitem{Gubser:2000nd}
S.~S. Gubser, {\it {Curvature singularities: The Good, the bad, and the
  naked}},  {\em Adv. Theor. Math. Phys.} {\bf 4} (2000) 679--745,
  [\href{http://xxx.lanl.gov/abs/hep-th/0002160}{{\tt hep-th/0002160}}].
  
\bibitem{Keeler:2013msa}
C.~Keeler, G.~Knodel, and J.~T. Liu, {\it {What do non-relativistic CFTs tell
  us about Lifshitz spacetimes?}},  {\em JHEP} {\bf 01} (2014) 062,
  [\href{http://xxx.lanl.gov/abs/1308.5689}{{\tt 1308.5689}}].

\bibitem{Ceresole:2007wx}
A.~Ceresole and G.~Dall'Agata, {\it {Flow Equations for Non-BPS Extremal Black
  Holes}},  {\em JHEP} {\bf 03} (2007) 110,
  [\href{http://xxx.lanl.gov/abs/hep-th/0702088}{{\tt hep-th/0702088}}].

\bibitem{Behrndt:2000ph}
K.~Behrndt and M.~Cvetic, {\it {Gauging of N=2 supergravity hypermultiplet and
  novel renormalization group flows}},  {\em Nucl. Phys.} {\bf B609} (2001)
  183--192, [\href{http://xxx.lanl.gov/abs/hep-th/0101007}{{\tt
  hep-th/0101007}}].

\bibitem{Bhattacharya:2012zu}
J.~Bhattacharya, S.~Cremonini, and A.~Sinkovics, {\it {On the IR completion of
  geometries with hyperscaling violation}},  {\em JHEP} {\bf 02} (2013) 147,
  [\href{http://xxx.lanl.gov/abs/1208.1752}{{\tt 1208.1752}}].

\bibitem{Knodel:2013fua}
G.~Knodel and J.~T. Liu, {\it {Higher derivative corrections to Lifshitz
  backgrounds}},  {\em JHEP} {\bf 10} (2013) 002,
  [\href{http://xxx.lanl.gov/abs/1305.3279}{{\tt 1305.3279}}].

\bibitem{Hristov:2009uj}
K.~Hristov, H.~Looyestijn, and S.~Vandoren, {\it {Maximally supersymmetric
  solutions of D=4 N=2 gauged supergravity}},  {\em JHEP} {\bf 11} (2009) 115,
  [\href{http://xxx.lanl.gov/abs/0909.1743}{{\tt 0909.1743}}].

\bibitem{Burda:2014jca}
P.~Burda, R.~Gregory, and S.~Ross, {\it {Lifshitz flows in IIB and dual field
  theories}},  {\em JHEP} {\bf 11} (2014) 073,
  [\href{http://xxx.lanl.gov/abs/1408.3271}{{\tt 1408.3271}}].

\bibitem{Donos:2010tu}
A.~Donos and J.~P. Gauntlett, {\it {Lifshitz Solutions of D=10 and D=11
  supergravity}},  {\em JHEP} {\bf 12} (2010) 002,
  [\href{http://xxx.lanl.gov/abs/1008.2062}{{\tt 1008.2062}}].

\bibitem{Hartnoll:2012wm}
S.~A. Hartnoll and E.~Shaghoulian, {\it {Spectral weight in holographic scaling
  geometries}},  {\em JHEP} {\bf 07} (2012) 078,
  [\href{http://xxx.lanl.gov/abs/1203.4236}{{\tt 1203.4236}}].

\bibitem{Donos:2012yi}
A.~Donos, J.~P. Gauntlett, and C.~Pantelidou, {\it {Semi-local quantum
  criticality in string/M-theory}},  {\em JHEP} {\bf 03} (2013) 103,
  [\href{http://xxx.lanl.gov/abs/1212.1462}{{\tt 1212.1462}}].

\bibitem{BarischDick:2012gj}
S.~Barisch-Dick, G.~L. Cardoso, M.~Haack, and S.~Nampuri, {\it {Extremal black
  brane solutions in five-dimensional gauged supergravity}},  {\em JHEP} {\bf
  02} (2013) 103, [\href{http://xxx.lanl.gov/abs/1211.0832}{{\tt 1211.0832}}].

\bibitem{Liu:2012wf}
J.~T. Liu and Z.~Zhao, {\it {Holographic Lifshitz flows and the null energy
  condition}},  \href{http://xxx.lanl.gov/abs/1206.1047}{{\tt 1206.1047}}.

\bibitem{Braviner:2011kz}
H.~Braviner, R.~Gregory, and S.~F. Ross, {\it {Flows involving Lifshitz
  solutions}},  {\em Class. Quant. Grav.} {\bf 28} (2011) 225028,
  [\href{http://xxx.lanl.gov/abs/1108.3067}{{\tt 1108.3067}}].

\bibitem{Donos:2010ax}
A.~Donos, J.~P. Gauntlett, N.~Kim, and O.~Varela, {\it {Wrapped M5-branes,
  consistent truncations and AdS/CMT}},  {\em JHEP} {\bf 12} (2010) 003,
  [\href{http://xxx.lanl.gov/abs/1009.3805}{{\tt 1009.3805}}].

\bibitem{Barisch-Dick:2013xga}
S.~Barisch-Dick, G.~L. Cardoso, M.~Haack, and A.~Veliz-Osorio, {\it {Quantum
  corrections to extremal black brane solutions}},  {\em JHEP} {\bf 02} (2014)
  105, [\href{http://xxx.lanl.gov/abs/1311.3136}{{\tt 1311.3136}}].

\bibitem{Hristov:2016vbm}
  K.~Hristov, S.~Katmadas and I.~Lodato,
  {\it {Higher derivative corrections to BPS black hole attractors in 4d gauged supergravity}},
   [\href{http://xxx.lanl.gov/abs/1603.00039}{{\tt 1603.00039}}].

\bibitem{workinprogress}
G.~L. Cardoso, K.~Goldstein, M.~Haack, and S.~Nampuri, {\it {work in progress}}.

\bibitem{Andrianopoli:1996vr}
L.~Andrianopoli, M.~Bertolini, A.~Ceresole, R.~D'Auria, S.~Ferrara, and
  P.~Fre', {\it {General matter coupled N=2 supergravity}},  {\em Nucl. Phys.}
  {\bf B476} (1996) 397--417,
  [\href{http://xxx.lanl.gov/abs/hep-th/9603004}{{\tt hep-th/9603004}}].

\bibitem{Freedman:2012zz}
D.~Z. Freedman and A.~Van~Proeyen, {\em {Supergravity}}.
\newblock Cambridge Univ. Press, Cambridge, UK, 2012.

\end{thebibliography}

\providecommand{\href}[2]{#2}\begingroup\raggedright\endgroup

\end{document}